# Robustness, Stability and Efficiency of Phage λ Gene Regulatory Network: Dynamical Structure Analysis

X.-M. Zhu[1)], L. Yin[2)], L. Hood[3)], and P. Ao[4),*]

[1)] GenMath, Corp. 5525 27th Ave.N.E., Seattle, WA 98105, USA
[2)] School of Physics, Peking University, Beijing 100871, PR China
[3)] Institute for Systems Biology, 1441 N. 34 St., Seattle, WA 98103, USA
[4)] Departments of Mechanical Engineering and Physics, University of Washington, Seattle, WA 98195, USA

## Summary

Based on the recently developed dynamical structure theory for complex networks and the seminal work of Shea and Ackers in the 1980's, we formulate a transparent and concise mathematical framework for the gene regulatory network controlling phage λ life cycles, which naturally includes the stochastic effect. The dynamical structure theory states that the dynamics of a complex network is determined by its four elementary components: The dissipation (analogous to degradation) and stochastic force, the driving force determined by a potential and the transverse force. The potential may be interpreted as a landscape for the phage development in terms of attractive basins, saddle points, peaks and valleys. The dissipation gives rise to the adaptivity of the phage in the landscape defined by the potential: The phage always has the tendency to approach the bottom of the nearby attractive basin. The stochastic fluctuation gives the phage the ability to search around the potential landscape by passing through saddle points.

With molecular parameters in our model fixed primarily by the experimental data on wild type phage and supplemented by data on one mutant, our calculated results on mutants agree quantitatively with the available experimental observations on other mutants for protein number, lysogenization frequency, and a lysis frequency in lysogen culture. The calculation reproduces the observed robustness of the phage λ genetic switch. This is the first mathematical description which successfully represents such a wide variety of major experimental phenomena. Specifically we find: 1) The explanation for both the stability and the efficiency of phage λ switch is the exponential dependence of saddle point crossing rate on potential barrier height, a result of the stochastic motion in a landscape; 2) The positive feedback of *cI* repressor gene transcription, enhanced by the CI dimer cooperative binding, is the key to the robustness of the phage λ genetic switch against mutations and parameter fluctuations.

**Running title:** modeling robustness of phage λ regulatory network

* Corresponding author. E-mail: aoping@u.washington.edu; fax: (206) 685-8047; ph: (206) 543-7837

## I. Introduction

Bacterophage λ is a virus that grows on a bacterium[1,2,3]. It is one of the simplest living organisms. The chromosome of phage λ consists a single DNA molecule wrapped in a protein coat. Upon infection of the host *E. coli* cell, the phage λ injects its chromosome inside the bacterium and leaves the protein coat outside. Inside the bacterium it chooses one of two modes of growth. It can direct the cell to produce new λ phage particles, resulting in the lysis of the cell. Or, it can establish dormant residency in the lysogenic state, integrating its genome into the DNA of its host and replicate as a part of the host chromosome. In these two different life cycles, different sets of phage genes are expressed as a result of molecular interactions. Realistic modeling of the robustness and stability of such a process has been remained a major challenge in biocomputation.

The developmental process and genetic control of phage λ has been perceived as a paradigm for the molecular mechanisms determining developmental pathways[4,5]. It is believed that analogous molecular interactions are likely to underlie many developmental processes[2,3]. One wishes to acquire deep understandings on the regulation of major biologic functions on molecular level through the study of the gene regulatory network of phage λ. One of these functions is the programming of the epigenetic states: The ways the phage decides if it is going to follow lysogenic or lytic growing state. Over past 5 decades, the extensive biological investigations have provided a fairly good qualitative picture in this respect. There is a plausible scenario to guide the understanding of the experimental observations.

The maintenance and operation of the genetic switch is another function performed by the gene regulatory network[2,3]. The phage growing in lysogenic state remains dormant unless it is provoked. Switching to lytic state happens when a signal is sent to activate RecA proteins which cleave CI monomer, sending the phage into lytic growth. It is observed that the phage λ switch is both highly stable and highly efficient. When the phage grows in lysogenic state, it remains dormant for many generations. Spontaneous induction happens less than once in a million cell divisions. Once the phage is exposed to an appropriate signal, it changes to lytic state at almost 100% rate. The stability and efficiency of the genetic switch in phage λ is still considered a mystery. We do not have a good understanding from either the biologic or the physical-chemical side.

There have been continuous mathematical and numerical activities on modeling phage λ. The rationale is rather straightforward: The biological functions should emerge as the systems property from the model based on the molecular mechanism of phage regulatory elements and their independently measured parameters. The elegant physical-chemical model formulated by Shea and Ackers[6] for gene regulation of phage λ has become the base for the later studies. However, soon afterwards, Reinitz and Vaisnys[7] pointed out that the inconsistency between the theoretical results and experimental data may suggest additional cooperativity. Arkin *et al.*[8] performed stochastic simulation on phage λ development for the decision of lysogeny in the early stage. Recently, Aurell and Sneppen[9] analyzed the robustness of phage λ genetic switch, using a method based

Onsager-Machlup functional[10], and found that their theoretical analysis could not reproduce the robustness of phage λ switch.

The coexistence of the switch stability and switching efficiency is an apparent dilemma for the following reasons. The lysogenic state is exceptionally stable. The fluctuations in the growth environment and the intrinsic fluctuation in the gene regulatory network do not accidentally flip the switch. Then when the phage is threatened, how can the switching process become so complete with so little outside intervention? The question about internal inconsistencies in these models naturally arises: Whether the easily operable induction, or highly efficient switching, in Shea and Ackers' work[6] is a result of sacrificing the robustness of the gene regulatory network. Phrasing differently, if a model were so constructed that it faithfully reproduces the observed robustness of the gene regulatory network, whether or not it would lose the efficiency of the switch. Undoubtedly a credible model of phage λ should reproduce the properties of robustness, stability and efficiency of the genetic switch simultaneously. From such a model we should also be able to calculate the observed quantities of phage development such as the protein numbers and lysogenization frequencies. Laying a foundation to build such a mathematical framework against the experimental data of phage λ is the primary goal of the present work.

Our procedure is first to build a minimal model for the gene regulatory network of phage λ to quantitatively reproduce as many experimental results as possible and to avoid any qualitative disagreement with experimental data. In particular, we require the calculated results from this model to agree with the experiments measuring the robustness of the genetic switch. After successfully obtaining such a model, we then proceed to study the relationships between the robustness, stability and efficiency of the genetic switch. We include the stochastic fluctuations of protein numbers in the model. It is necessary to include such fluctuations not only because of their ubiquitous presence, also because they are the key quantities influencing the intrinsic stability of the genetic switch and switching efficiency.

By combining a newly developed powerful nonlinear dynamics analysis method which takes the stochastic force into account[11,12] and the previously established physical-chemical model[6] we formulate a mathematical framework to calculate the following quantities of epigenetic states and developmental paths: the protein numbers, the protein distributions, the lifetime of each state, and the lysogenization frequencies of mutants using the wild type as reference. We find that our calculated results agree quantitatively with the available experimental data.

Our method to solve the mathematical model has a unique advantage. It helps to bring a hidden layer of gene regulatory network structure up to surface. In the mathematical analysis, we calculate the potential for the gene regulatory network as a function of protein numbers. The potential may be interpreted as the landscape map of the phage development in terms of attractive basins, saddle points, peaks and valleys. Such a concept of landscape for development and differentiation had already been conceived long time ago in biology[13]. The phage evolving in this potential landscape resembles a charged particle moving in an electromagnetic field. Similar to the case in electrodynamics, the potential landscape provides a visualization of the dynamical structure of the gene regulatory network. The minima of the potential give the location of the possible epigenetic states. The height of the saddle point between the potential minima gives the timescale separating these epigenetic states.

With the workable model a new light on the biologic operation of the gene regulatory network from mathematical modeling requires can be shed by organizing the gene elements and by exploring their relationships in ways that can be easily done mathematically but cannot or have not been done experimentally. By such a manipulation, we may gain insight into the biological structure of the gene regulatory network. The encouragement comes from the fact that indeed our modeling demonstrates, for the first time, the desired robustness of the genetic switch. By mathematically eliminating the feedback mechanisms and then putting them back one by one, adjusting their relative strength, we find that the positive feedback of CI production, enhanced by the cooperative bindings of CI dimers on the operators, appears to be the key to the robustness of the switch structure.

Our formulation provides an intuitively appealing solution to the stability-efficient dilemma. The mathematical equivalence of a genetic switch is a charged particle moving in a double-well potential subject to a random force. The lifetime in a potential minimum, representing the lifetime of an epigenetic state, has an exponential dependence on the potential barrier height. By lowing the potential barrier height, the lifetime of phage in lysogenic state decreases drastically. The phage evolves quickly towards to more favorable growing state, the lytic state. It is this exponential dependence that is responsible for both efficiency in switching and stability of epigenetic state.

We will also discuss the practical relevancy of our formulation. For any newly introduced theoretical quantity, one of the crucial questions to ask is how to probe its structure experimentally. An analogy may be drawn from the concept of pressure in physics and chemistry. Besides its thermodynamic significance and its microscopic origin, pressure is also a quantity that can be directly measured macroscopically. We demonstrate that the potential and the friction of gene regulatory network introduced in the present work also have such transparent connections with experiments. One of the main problems in determining the dynamical structure from experiments, the so-called ‘reverse engineering’, is the inability to uniquely determine the microscopic dynamical parameters. Our method may provide a needed tool to solving this problem by designing more relevant experiments to explore the relationship of measurable quantities at the given description level.

The rest of the paper is organized as follows. Salient biological experimental studies are summarized in section II. Key previous biochemical modeling elements are summarized in section III. The purpose of section II and III is to provide a concise and coherent biological, chemical, and physical background for our modeling. The dynamical structure analysis method is discussed in section IV within the minimal model of the phage $\lambda$.

Calculated results, the comparison to biological data, and discussions are presented in detail in section V. In section VI the present work is put into perspective.

## II. λ Switch

The genetic switch controlling and maintaining the function of phage λ consists of two regulatory genes, *cI* and *cro*, and the regulatory regions, $O_R$ and $O_L$ on the λ DNA. Established lysogeny is maintained by the protein CI which blocks operators $O_R$ and $O_L$, preventing transcription of all lytic genes including *cro*[2,3]. In lysogeny the CI number functions as an indicator of the state of the bacterium: If DNA is damaged such as by the UV light the protease activity of RecA is activated, leading to degradation of CI. A small CI number allows for transcription of the lytic genes, starting with *cro*, the product of which is the protein Cro.

The decision making, or the switching, is centered around operator $O_R$, and consists of three binding sites $O_{R1}$, $O_{R2}$ and $O_{R3}$, each of which can be occupied by either a Cro dimer or a CI dimer[2,3]. As illustrated in Fig.1, these three binding sites control the activity of two promoters $P_{RM}$ and $P_R$ for respectively *cI* and *cro* transcriptions. The transcription of *cro* starts at $P_R$, which partly overlaps $O_{R1}$ and $O_{R2}$. The transcription of *cI* starts at $P_{RM}$, which overlaps $O_{R3}$. The affinity of RNA polymerase for the two promoters, and subsequent production of the two proteins, depends on the pattern of Cro and CI bound to the three operator sites and thereby establishes lysogeny with about 500 CI molecules per bacterium. If, however, CI number becomes sufficiently small, the increased production of Cro flips the switch to lysis.

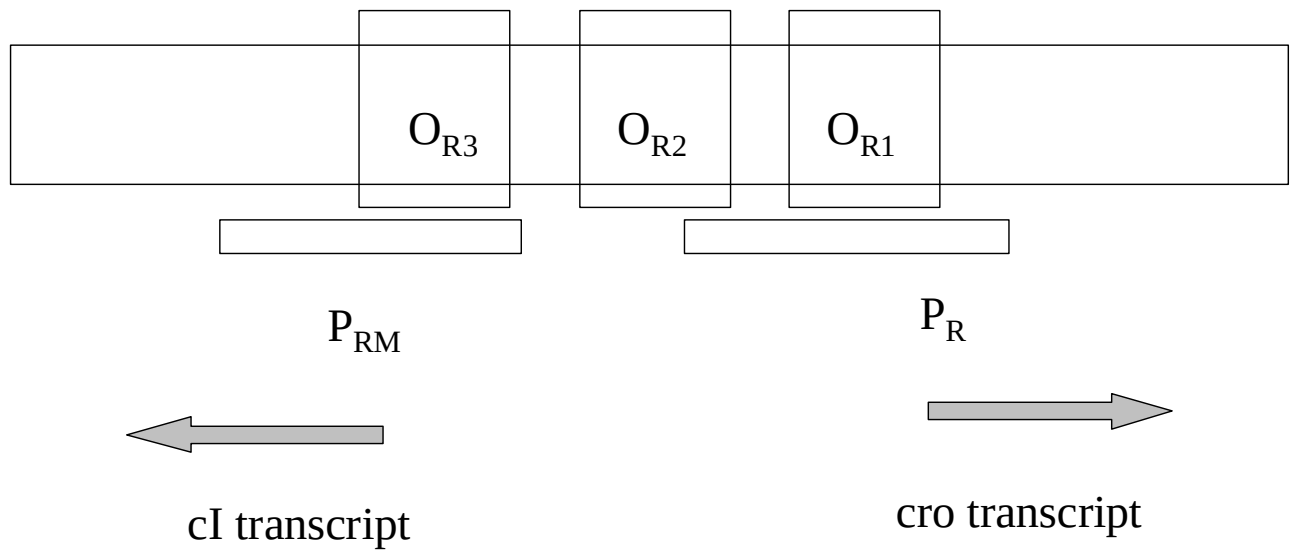

Fig. 1. The $O_R$ controlling region of the phage λ gene regulatory network. Our mathematical study indicates that the cooperative binding of CI dimers at the $O_{R1}$ and $O_{R2}$ sites is a key to the robustness of the gene regulatory network. Such a cooperative binding enhances the positive CI feedback. When the CI positive feedback is turned on by the existing CI dimers, CI proteins are transcribed. The phage evolves to the lysogenic state. Otherwise Cro proteins are transcribed and the phage evolves to lytic state.

Quantitative experimental study of the stability in the switching of bacteriophage λ has a long history. Recently, the frequency of spontaneous induction in strains deleted for the *recA* gene has been reported independently by three groups[14,15,16], which was reviewed by Aurell *et al.*[16]. They all confirmed two earlier fundamental observations that there is a switching behavior and that the switch is stable. In addition, they all obtained consistent numerical values for the switching frequency, in spite of the use of different strain backgrounds, done at different continents and at different times. Computational attempts to quantitatively understand this behavior have not been successful, even permitting the possibility that the wild type may be more stable[9,16].

More recent data[17] suggest that the wild type may be two orders of magnitude more stable than previously observed value in Ref.[15]: The switch rate to lytic state may be less than $4\times10^{-9}$ per minute. In addition to the call for more experimental studies, this puts the theoretical modeling in a more challenging position. In the following we will use this new wild type data as our main input to completely fix our model. We will also discuss the previous data to illustrate a pronounced exponential sensitivity in our modeling.

## III. Physical Chemical Model

The CI and Cro protein molecules in the cell are assumed to be in homeostatic equilibrium. There are not always the same numbers of CI and Cro dimers bound to the operators at any particular time. These numbers are fluctuating, and the equilibrium assumption should give the size of these fluctuations. The key inputs are CI and Cro dimerization constants and the Gibbs free energies for their bindings to the three operator sites $O_{R1}$, $O_{R2}$ and $O_{R3}$[18,19,20,21,22,23,24].

Following Ackers *et al.*[25] and Aurell *et al.*[16], we encode a state s of CI and/or Cro bound to $O_R$ by three numbers (i,j,k) referring to $O_{R3}$, $O_{R2}$ and $O_{R1}$ respectively. The coding is 0 if the corresponding site is free, 1 if the site is occupied by a CI dimer, and 2 if the site is occupied by a Cro dimer. The probability of a state s with $i_s$ CI dimers and $j_s$ Cro dimers bound to $O_R$ is in the grand canonical approach of Shea and Ackers[6]

$$p_R(s) = Z^{-1} [CI]^{i_s} [Cro]^{j_s} [RNAp]^{k_s} \exp(-\Delta G(s)/T) \qquad \text{Eq.(1)}$$

For example, if CI occupies $O_{R1}$, Cro $O_{R2}$ and $O_{R3}$, we have $i_s = 1$, $j_s = 2$, $k_s = 0$, and $p_R(s) = p_R(221)$. There are total 40 states represented by s. The normalization constant $Z$ is determined by summing over $s$: $Z = \sum_s [CI]^{i_s} [Cro]^{j_s} [RNAp]^{k_s} \exp(-\Delta G(s)/T)$ . Here [ ] denotes the corresponding protein dimer concentration in the bacterium.

RNA polymerase (RNAp) occupies either $O_{R1}$ and $O_{R2}$, or $O_{R2}$ and $O_{R3}$, not other configurations. We further simplify the expression of $p_R(s)$ by noticing that the controlling of $O_R$ is operated by CI and Cro proteins, not RNAp[2,3]. If $O_{R1}$ and $O_{R2}$ are unoccupied by either CI or Cro, RNAp binds to them with a probability determined by RNAp binding energy. The case that RNAp first binds to $O_{R1}$ and $O_{R2}$, then blocking CI and Cro binding is excluded based on the assumption that only CI and Cro controls the regulatory behavior. In addition to experimental observation, this assumption is justifiable if the time scale associated with CI and Cro binding is shorter than the RNAp binding. Except for an overall constant, which we include into the rate of transcription, the RNAp binding is no longer relevant. We therefore take it out of the expression $p_R(s)$. The total number of states is reduced to 27. This simplification was first used by Aurell and Sneppen[9]. We will drop the subscript R for binding probability $p_R$. We should point out that previous experimental and theoretical results had been concisely reviewed by Aurell *et al.*[16], whose convention we shall follow.

The dimer and monomer concentrations are determined by the formation and de-association of dimers, which gives the relation of dimer concentration to the total concentraion of proteins as

$$[CI] = [N_{CI}]/2 + \exp(\Delta G_{CI}/RT)/8 + ([N_{CI}]\exp(\Delta G_{CI}/RT)/8 + \exp(2\Delta G_{CI}/RT)/64)^{1/2} \qquad \text{Eq.(2)}$$

Here $\Delta G_{CI}$ = - 11.1 kcal/mol is the dimer association free energy for CI.
Similar expression for [Cro] is

$$[Cro] = [N_{Cro}]/2 + \exp(\Delta G_{cro}/RT)/8 + ([N_{Cro}]\exp(\Delta G_{Cro}T)/8 + \exp(2\Delta G_{Cro}T)/64)^{1/2} \qquad \text{Eq.(3)}$$

Here $\Delta G_{Cro}$ = - 7 kcal/mol is the dimer association free energy for Cro. Here $[N_{CI}]$ and $[N_{Cro}]$ are the monomer concentrations of CI and Cro respectively.

CI and Cro are produced from mRNA transcripts of cI and cro, which are initiated from promotor sites $P_{RM}$ and $P_R$. The rate of transcription initiation from $P_{RM}$ when stimulated by CI bound to $O_{R2}$ is denoted $T_{RM}$, and when not stimulated $T_{RM}^u$. The number of CI molecules produced per transcript is $E_{cI}$. The overall expected rate of CI production is

$$f_{CI}(N_{CI}, N_{Cro}) = T_{RM} E_{cI} [\, p(010)+ p(011)+ p(012)\,] + T_{RM}^u E_{cI} [\, p(000) + p(001)+ p(002)+ p(020)+ p(021) + p(022)\,] . \qquad \text{Eq.(4)}$$

Here $N_{CI}$ and $N_{Cro}$ are the protein numbers for CI and Cro inside the bacterium respectively. The converting factor between the protein concentration and the corresponding protein inside the bacterium is listed in Table I. Similarly, the overall expected rate of Cro production is

$$f_{Cro}(N_{CI}, N_{Cro}) = T_R E_{cro} [p(000)+ p(100)+ p(200)]. \qquad \text{Eq.(5)}$$

We use $T_{RM}$, $E_{cI}$, $E_{cro}$, and $T_{RM}^u$ from Aurell and Sneppen[9], which were deduced from the resulting protein numbers in lysogenic and lytic states.

The free energies $\Delta G(s)$ are determined from *in vitro* studies. The *in vivo* conditions could be different. The measured protein-DNA affinities could depend sensitively on the ions present in the buffer solutions as well as other factors. This observation will be important in our comparison between theoretical results and experimental data. On the other hand, the *in vivo* effects of such changes should be compensated for, as e.g. changed KCl concentrations are by putrescine[26] and other ions and crowding effects[27].

We note that Record *et al.*[27] already observed that there may exist a significant difference between *in vivo* and *in vitro* molecular parameters. The data quoted in Darling *et al.* was obtained at KCl concentration of 200mM, which resembles *in vivo* conditions. Therefore, though we expect a difference between the *in vivo* and *in vitro* data, the difference may not be large.

The mathematical model which describes the genetic regulation in Fig.1 is a set of coupled equations for the time rate of change of numbers of CI and Cro in a cell[7]:

$$dN_{CI}(t)/dt = F_{CI}(N_{CI}(t), N_{Cro}(t))$$
$$dN_{Cro}(t)/dt = F_{Cro}(N_{CI}(t), N_{Cro}(t)) \qquad \text{Eq.(6)}$$

where the net production rates are

$$F_{CI} = f_{CI}(N_{CI}, N_{Cro}) - N_{CI}/\tau_{CI}$$
$$F_{Cro} = f_{Cro}(N_{CI}, N_{Cro}) - N_{Cro}/\tau_{Cro} \qquad \text{Eq. (7)}$$

Here dN/dt is the rate N changes. The production terms $f_{CI}$ and $f_{Cro}$ are functions of CI and Cro numbers in the bacterium. With no Cro in the system, the curve of $f_{CI}$ *vs*. CI number has been experimentally measured[28]. As reviewed in Aurell *et al.*[16] these measurements are consistent with the best available data on protein-DNA affinities[18,20,29] dimerization constants[30], initiation rates of transcriptions of the genes, and the efficiency of translation of the mRNA transcripts into protein molecules. The decay constant $\tau_{CI}$ is proportional to the bacterial life-time, since CI molecules are not actively degraded in lysogeny, while $\tau_{Cro}$ is about 30% smaller[31]. We comment that there is considerably more experimental uncertainty in the binding of Cro, both to other Cro and to DNA, than the binding of CI, see e.g. Darling *et al.*[23,24]. As a minimal model of the switch, we take $\tau_{CI}$ and $\tau_{Cro}$ from data, and deduce $f_{CI}$ and $f_{Cro}$ at non-zero number of both CI and Cro with a standard set of assumed values of all binding constants, which are summarized by Aurell *et al.*[16] and are adopted here (Table I with differences in cell volume and converting factor, as well as the *in vivo* and *in vitro* differences)

## IV. Stochastic Effect and Dynamical Structure Analysis

### IV.1 Stochastic dynamics

If the numbers of CI and Cro were macroscopically large, then Eq.(1) would be an entirely accurate description of the dynamics, because the fluctuation in number is an order of $N^{1/2}$ and the correction is an order of $1/N^{1/2}$. The numbers are however only in the range of hundreds. Hence the fluctuation is not negligible. The actual protein production process is influenced by many chance events, such as the time it takes for a CI or a Cro in solution to find a free operator site, or the time it takes a RNA polymerase molecule to find and attach itself to an available promoter, suggesting more stochastic sources. As a minimal model of the network with finite-N noise, we therefore consider the following system of two coupled stochastic differential equations, with two independent standard Gaussian and white noise sources:

$$dN_{CI}/dt = F_{CI} + \zeta_{CI}(t)$$
$$dN_{Cro}/dt = F_{Cro} + \zeta_{Cro}(t) \qquad \text{Eq.(8)}$$

In case integration and differentiation are involved, the Ito calculus will be adopted. We further assume that the means of the noise terms are zero, *i.e.* $\langle\zeta_{CI}(t)\rangle = \langle\zeta_{Cro}(t)\rangle = 0$, with the variance

$$\langle \zeta_{CI}(t)\, \zeta_{CI}(t') \rangle = 2\, D_{CI}\, \delta(t-t')$$
$$\langle \zeta_{Cro}(t) \zeta_{Cro}(t') \rangle = 2\, D_{Cro}\, \delta(t-t')$$
$$\langle\zeta_{CI}(t)\, \zeta_{Cro}(t') \rangle = 0 \qquad \text{Eq.(9)}$$

Eq.(9) defines a 2×2 diffusion matrix $D$. The noise strength may contain contributions from the production and decay rates, assuming each is dominated by one single independent reaction, as used by Aurell and Sneppen[9]. Such a noise may be called the `intrinsic' noise. Other noise sources, 'extrinsic' noises, also exist[32]. We treat the noise to incorporate both intrinsic and extrinsic sources: All are assumed to be Gaussian and white. The consistent of this assumption should be tested experimentally, as will be the case below. Certain probability events, however, may not behave as Gaussian and white in the context of modeling, which can be determined by separate biological experiments, such as the $p_{RM}$240 mutation[17] to be discussed below.

It has been demonstrated[11,12] that there exists a unique decomposition such that the stochastic differential equation, Eq.(8), can be transformed into the following form:

$$[\Lambda(\mathbf{N}) + \Omega(\mathbf{N})]\, d\mathbf{N}/dt = -\nabla U(\mathbf{N}) + \xi(t) \qquad \text{Eq.(10)}$$

with the semi-positive definite symmetric 2×2 matrix $\Lambda$, the anti-symmetric 2×2 matrix $\Omega$, the single valued function U, and the two dimensional vectors:

$$\mathbf{N}^{\tau} = (N_{CI}, N_{Cro});$$
$$\nabla = (\partial / \partial N_{CI},\ \partial / \partial N_{Cro});$$
$$\xi^{\tau} = (\xi_{CI}, \xi_{Cro}), \qquad \text{Eq.(11)}$$

here $\tau$ means the transpose of the vector. The connection between the noise $\xi$ and the matrix $\Lambda$ is similar to that of $\zeta$ and $D$ of Eq.(9):

$$\langle\xi(t)\rangle = 0$$
$$\langle \xi_{CI}(t)\, \xi_{CI}(t') \rangle = 2\, \Lambda_{CI}\, \delta(t-t')$$
$$\langle \xi_{Cro}(t)\, \xi_{Cro}(t') \rangle = 2\, \Lambda_{Cro}\, \delta(t-t')$$
$$\langle\xi_{CI}(t)\, \xi_{Cro}(t') \rangle = 0 \qquad \text{Eq.(12)}$$

The decomposition from Eq.(8 ) and Eq.(9 ) to Eq.(10) and Eq.(12) is determined by the following set of equations:

$$\nabla\times[(\Lambda + \Omega)\,\mathbf{F}] = 0 \qquad \text{Eq.(13)}$$

$$(\Lambda + \Omega)D(\Lambda - \Omega) = \Lambda\,. \qquad \text{Eq.(14)}$$

One may solve for $\Lambda$, $\Omega$ in terms of **F** and $D$ from Eq.(13) and Eq.(14). Indeed, this can be formally done. Once $\Lambda$, $\Omega$ are known, the requirement that Eq.(10) can be reduced to Eq.(8) gives $(\Lambda + \Omega)\,\mathbf{F} = -\nabla U(\mathbf{N})$, which is used to obtain U. In general this decomposition is an involved mathematical and numerical endeavor. Further simplification follows from the simplification of friction matrix. Typically the diffusion matrix $D$ is unknown biologically: There are no enough measurements to fix the noise explicitly. Therefore we may treat the semi-positive definite symmetric matrix $\Lambda$ as parameters to be determined experimentally. In our calculation, we then assume that $D$ is a diagonal matrix. Following from Eq.(14), $\Lambda$ is a diagonal matrix for two dimensional case. The experimentally measured fraction of $recA^{-1}$ lysogens that have switched to lytic state is used to determine the elements of $\Lambda$.

Here we would like to give an intuitive interpretation of the mathematical procedure. Eq.(8) corresponds to the dynamics of a fictitious massless particle moving in two dimensional space formed by the two protein numbers $N_{CI}$ and $N_{Cro}$, with both deterministic and random forces. It is easy to check that in general $\nabla\times\mathbf{F}(\mathbf{r}) \neq 0$ and $\nabla\bullet\mathbf{F}(\mathbf{r}) \neq 0$. Therefore $\mathbf{F}(\mathbf{r})$ cannot be simply represented by the gradient of a scalar potential due to both the force transverse to the direction of motion and force of friction. The simplest case in two dimensional motion when both transverse force and friction exist is an electrically charged particle moving in the presence of both magnetic and electric fields, which is precisely in the form of Eq.(10).

Proceeding from Eq.(10), we note that we may interpret the semi-positive definite symmetric $\Lambda$ matrix as the friction matrix, and the antisymmetric matrix $\Omega$ as the result of a `magnetic' field. The friction matrix represents the dissipation in physics. It is analogous to the degradation in biology. The scalar function U takes the role of a potential function which would determine the final steady distribution of the phage. The global equilibrium will be reached when the final distribution function is given by

$$\rho(N_{CI},N_{Cro}) = \exp(-U(\mathbf{N}))/\int dN_{CI} \int dN_{Cro} \exp(-U) \qquad \text{Eq.(15).}$$

The potential U, the landscape of the system, is depicted in Fig. 2 (*c.f.* Fig. 4).

The phage sees two minima and one saddle point in the potential landscape. Those two minima correspond to the lytic and lysogenic states. Once the phage is at one of the minimum, the probability rate for it to move into another minimum is given by the Kramers rate formulae in the form[33,34]:

$$P = \omega_0 \exp(-\Delta U_b) \qquad \text{Eq.(16)}$$

with the potential barrier height $\Delta U_b = (U_{saddle} - U_{initial\ minimum})$, the difference in potential between the saddle point and the initial minimum, and the time scale, the attempt frequency $\omega_{0,}$ determined by the friction, the curvatures around the saddle. We remark here that the attempt frequency is in general a complicated function of dynamical quantities in Eq,(10). It's form will be determined empirically in the present paper. We refer readers to Ref.[34] for the general mathematical discussions.

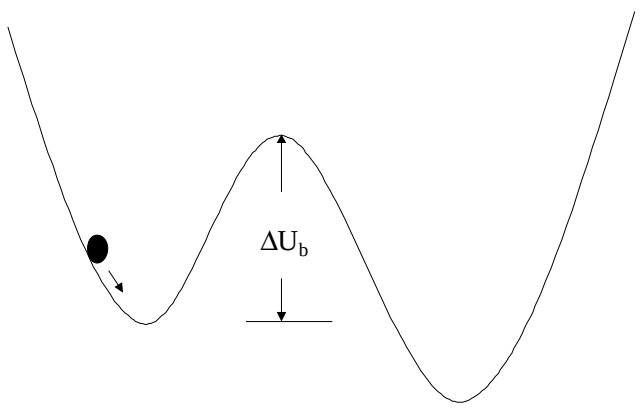

Fig. 2. Illustration of the dynamical structure of a gene regulatory network. The dynamic state of the network is represented by a particle whose position is given by instantaneous protein numbers. The potential function maps a landscape in the protein number space. For a genetic switch, there are two potential minima corresponding to two epigenetic states. The area around each of the minima forms the attractive basin. The state of the network always tends to relax to one of the minima. The fluctuation may bring the network from one minimum to another with a rate given by Kramers rate formulae[33,34].

### IV.2 Dynamical structure of the gene regulatory network

Eq.(10) gives the dynamical structure of the gene regulatory network in terms of its four components: the friction, the potential gradient of the driving force, the transverse force, and the stochastic force. Such a dynamical structural classification serves two main purposes. It provides a concise description for the main features of the gene regulatory network by itself and it provides a quantitative measure to compare different gene regulatory networks, for instance between the wild phage and its mutant.

The potential may be interpreted as the landscape map of the phage development. Each of the epigenetic state is represented by a potential minimum and its surrounding area forms an attractive basin. The dissipation represented by the friction gives rise to the adaptivity of the phage in the landscape defined by the potential: The phage always has the tendency to approach the bottom of the nearby attractive basin. The potential change near the minimum, together with the friction, gives the time scale of relaxation: The time it takes to reach equilibrium after the epigenetic state is perturbed. Once we know the friction and the potential around the minimum, we have a good grasp of the relaxation time, $\tau = \eta/U''$, here $\eta$ is the strength of friction, $U''$ the second derivative of potential, both in one dimensional approximation along a relevant axis. The relaxation time is independent of the amplitude of the perturbation near the potential minimum, when $U''$ is a constant.

Two remarks are in order here. 1. The meaning of friction matrix is the same as in mechanics: If there is no external driving force, the system tends to stop at its nearby minimal position. The closest corresponding concept in biology is `degradation': There is always a natural protein state under given condition. 2. It turns out that the transverse force is not a dominant factor in the present switch-like behavior. However, its existence is the necessary condition for oscillatory biological behaviors, which will not be discussed further in the present paper.

Another time scale provided by the potential is the lifetime of the epigenetic state, which is given by the Kramers rate formulae, Eq. (16), through the potential barrier height. Such a scale measures the stability of epigenetic state in the presence of fluctuating environment. In the case of phage λ, the lifetime for lysogenic state is very long, unless the phage is mutated at its operator sites. When the phage is provoked, the height of the potential barrier separating lysogenic and lytic states is reduced. The lifetime of lysogenic state is drastically reduced due to its exponential dependence on the barrier height and switching takes place. Looking at it from a different angle, the stochastic force gives the phage ability to search around the potential landscape by passing through saddle points and drives the switching event. The Kramer rate formulae is a quantitative measure of this optimization ability.

## V. Results

### V. 1 Determining *in vivo* parameters.

First we need to decide the free energies to be used in the theoretical model. Without exception, all the binding energies measured so far for phage λ are determined from *in vitro* studies. The difference between the *in vivo* condition and the *in vitro* condition could include the ion concentration in the buffer solutions and the spatial configuration of the genomic DNA, for instant looping[35,36,37]. The relative large change of the cooperative energy from *in vitro* to *in vivo* in Table I may be partly due to the looping effect, though there is no direct consideration of looping in present model. We note that in the *in vivo* conditions all the operators are in the same kind of environment, including the ion condition and the DNA configuration. The reason for the latter is that the operators are closely located to each other in the genome. If there is a bending of the genomic DNA which increase or decrease DNA-protein bindings, these closely located and short operator sites are mostly likely experience the same amount of change. Therefore, we assume that in addition to the *in vitro* DNA-protein binding energy, overall binding energy differences are added to all the CI and Cro protein respectively:

*in vivo* binding energy for CI (Cro) = *in vivo* binding energy for CI + $\Delta G_{CI}$ ($\Delta G_{Cro}$).

To determine $\Delta G_{CI}$($\Delta G_{Cro}$), we need more experimental input than the *in vitro* measurement. To avoid unnecessary uncertainly, in the model we build, we try to include the minimal number of parameters. The cooperative binding between two CI dimers is included. The cooperative bindings between two Cro dimers, between CI and Cro dimers, the unspecific CI and Cro bindings are not included. Our later calculation verifies that CI cooperative binding is essential to the genetic switch properties while the bindings we ignore do not have significant influence on the calculated results. There are three parameters we need to adjust, the difference between *in vivo* and *in vivo* binding energy for CI ($\Delta G_{CI}$), for Cro ($\Delta G_{Cro}$), and for the cooperativity of CI dimers ($\Delta G$(cooperative)).

We first use the CI numbers of both wild type and mutant $\lambda O_R121$ to determine $\Delta G_{CI}$, then we determine ΔG(cooperative) and $\Delta G_{Cro}$ by requiring both the lytic and lysogenic states of wild type are equally stable, calculated from Kramers rate formulae. The adjusted *in vivo* binding energies and other parameters we use for the modeling are given in Table I. Using these adjusted parameters the robustness of the phage's genetic switch is reproduced (shown in Fig. 3).

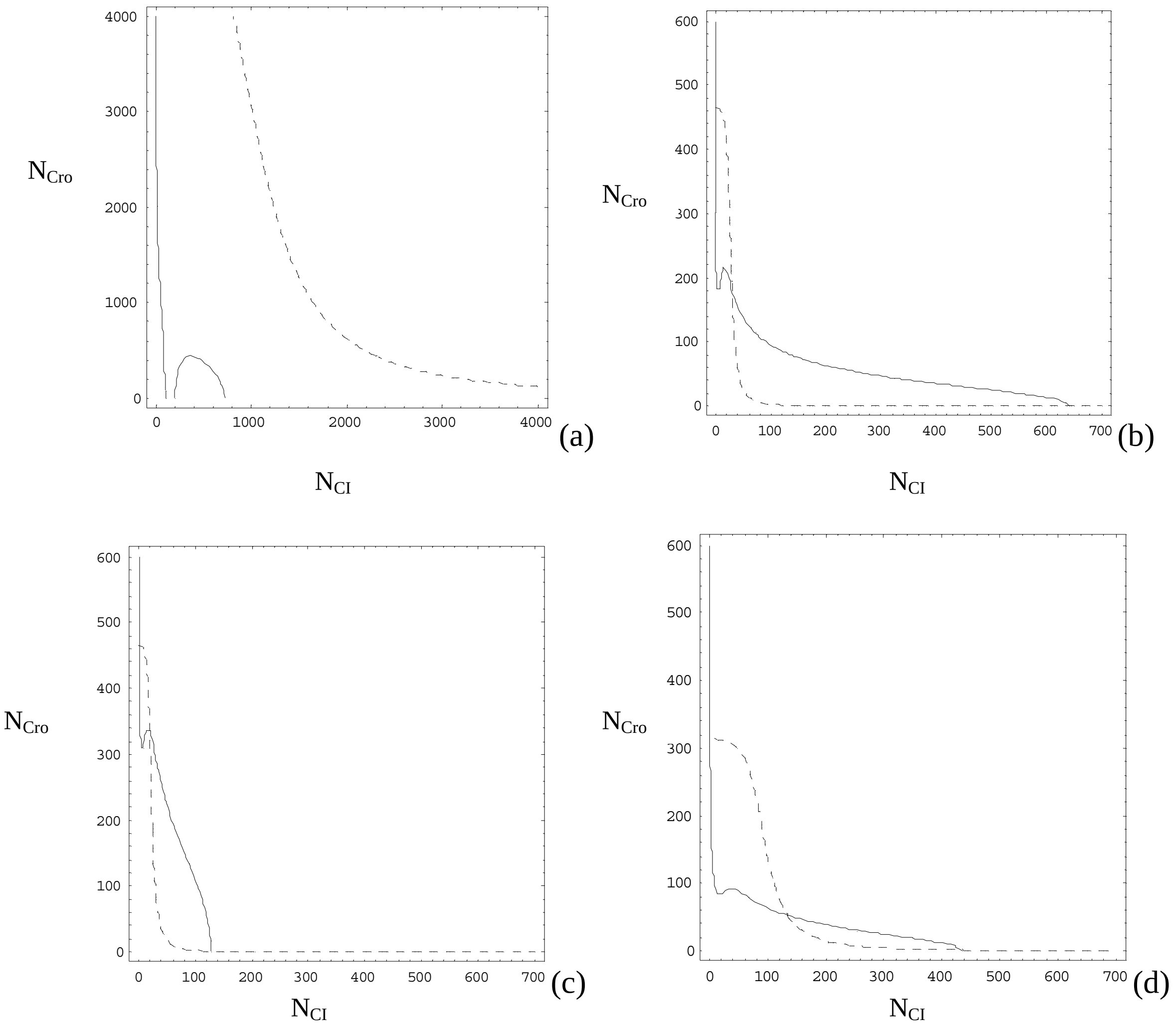

Fig.3. Lines of $d\langle N_{CI}\rangle/dt = 0$ (solid) and $d\langle N_{Cro}\rangle/dt = 0$ (dashed), here $\langle\ \rangle$ is the average to stochastic force, for: (a) the wild type phage, $\lambda O_R321$, with parameters taken directly from *in vitro* measurement; (b) the wild type phage with parameters adjusted allowing *in vivo* and *in vitro* differences; (c) mutant $\lambda O_R121$ with the parameters used in (b); (d) mutant $\lambda O_R323$ with the parameters used in (b). For (b), (c) and (d), these two lines have three intersections. These three fixed point in Eq.(2) coincide with the potential extrema, minima and saddle point, in Eq.(10). The mutant $\lambda O_R123$, not plotted here, has only one fixed point corresponding to lytic growth.

The mutant $\lambda O_R3'23'$ studied by Little *et al.*[15] was characterized by Hochschild *et al.*[38] for binding to $O_R3'$. To produce the desired protein level, we find that the binding energy between $O_R3'$ and Cro protein is 1.8 kcal/mol smaller than that of the $O_R3$ and Cro protein, consistent with the result of Hochschild *et al*. The CI binding energy from $O_R3$ to $O_R3'$ is slightly increased, 1 kcal/mol, also consistent with the measurement.

| | |
|---|---|
| RT<br>Effective bacterial volume<br>$E_{cro}$<br>$E_{cI}$<br>$T_{RM}$<br>$T_{RM}^u$<br>$T_R$<br>$\tau_{CI}$<br>$\tau_{Cro}$<br>Converting factor between protein number and concentration | 0.617 kcal/mol<br>$0.7\times10^{-15}$ *l*<br>20<br>1<br>0.115 /s<br>0.01045 /s<br>0.30/s<br>2943 s<br>5194 s<br>$1.5\times10^{-11}$ |
| *in vitro* free energy differences for wild type λ<br>ΔG (001)<br>ΔG (010)<br>ΔG (100)<br>ΔG (011)<br>ΔG (110)<br>ΔG (111)<br>ΔG (002)<br>ΔG (020)<br>ΔG (200)<br>ΔG(cooperative)<br>dimerization energy<br>$\Delta G_{CI2}$<br>$\Delta G_{Cro2}$ | <br>-12.5 kcal/mol<br>-10.5 kcal/mol<br>-9.5 kcal/mol<br>-25.7 kcal/mol<br>-22.0 kcal/mol<br>-35.4 kcal/mol<br>-14.4 kcal/mol<br>-13.1 kcal/mol<br>-15.5 kcal/mol<br>-2.7 kcal/mol<br><br>-11.1 kcal/mol<br>-7.0 kcal/mol |
| *in vitro* free energy differences for $O_R3'$ binding<br>ΔG (100)<br>ΔG (200) | <br>-10.5 kcal/mol<br>-13.7 kcal/mol |
| *in vivo* free energy differences - *in vitro* free energy differences<br>$\Delta G_{CI}$<br>$\Delta G_{Cro}$<br>ΔG(cooperative) | <br>–2.5 kcal/mol<br>-4.0 kcal/mol<br>-3.7 kcal/mol |

Table I. Parameters used in the modeling. CI dimer affinities to $O_{R1}$, $O_{R2}$ and $O_{R3}$ are after Darling *et al.*[23,24]. Cro dimer affinities to $O_{R1}$, $O_{R2}$ and $O_{R3}$ are after Takeda *et al.*[19,29], Jana *et al.*[22], Kim *et al.*[20], as stated in Aurell *et al.*[16]. The CI dimerization energy is from Koblan and Ackers[30], Cro dimerization energy from Jana *et al.*[21,22]. The bacterial volume is taken from Bremmer and Dennis[39]. $E_{CI}$ and $E_{Cro}$ are taken from Shean and Gottesman[40], Ringquist *et al.*[41], and Kennell and Riezman[42]. The *in vitro* parameters have been summarized by Aurell *et al.*[16], which we largely follow. However, we here point out two differences: 1) Our effective bacterial volume, estimating from the typical size of the bacterium, is about factor 3 smaller; and 2) The normalization factor for the concentrations, calculated against the numbers of water molecules, is about factor of 60 smaller. The difference between *in vivo* and *in vitro* values is consistent with qualitative experimental observation[2]. The wild type data of Little[17] is used to determine the *in vivo* and *in vitro* difference.

We assume that friction matrix $\Lambda$ is a diagonal constant matrix. Similar to Aurell and Sneppen[9], we assume the stochastic fluctuations in Eq.(2) scale with the square root of protein number divided by relaxation time: $D_{CI}$ = Const$\times\tau_{CI}/N_{CI}$ (lysogen), and $D_{Cro}$ = Const$\times\tau_{Cro}/N_{Cro}$ (lysis), where $N_{CI}$ (lysogen) is the CI number at lysogenic state and $N_{Cro}$ (lysis) is the Cro number at the lytic state. The Const is to be determined by experiments. In Eq.(7), we note that if the antisymmetric matrix $\Omega$ is small, then $\Lambda$ is the inverse of $D$. We calculate $\Omega$ assuming $\Lambda = D^{-1}$ and find that indeed in the regions of concern, i.e. the potential valley connecting two potential minima through the saddle points, $A$ is negligible. The final parameters we have used are

$$\Lambda_{11} = 0.056\times\tau_{CI}/N_{CI}(\text{lysogen})$$
$$\Lambda_{22} = 0.040\times\tau_{Cro}/N_{Cro}(\text{lysis})\ . \qquad \text{Eq.(17)}$$

## V.2 Dynamical structure of the phage λ gene regulatory network

The original problem, described by Eq.(8), may be interpreted as a set of two dimensional differential equation describing a particle motion, if we view the protein number $N_{CI}$ and $N_{Cro}$ as the coordinates and the particle position to be ($N_{CI}(t)$ $N_{Cro}(t)$) at time t. There is a deterministic force $\mathbf{F}^{\tau}$ = ($F_{CI}$, $F_{Cro}$) and a stochastic force acting on such a particle. The deterministic force has the characteristics of a friction, a potential force, and a transverse force at the same time. The decomposition we have discussed earlier, Eq.(10) allows to separate these components. We discuss them respectively here.

The wild type phage λ and some of its mutants sees two minima and one saddle point in the potential energy landscape (Fig. 4). Those two minima correspond to the lytic and lysogenic states (*cf*. Fig. 2). The positions of the potential minima give the average protein number for lytic and lysogenic states. There is a relatively narrow valley connecting these two minima. The highest point along this valley is the saddle point. Since the areas with large potential are not easily accessible and the low-lying potential region forms a valley, we may visualize the potential along the valley and illustrate it in a one dimensional graph as shown in Fig.2 and Fig.12.

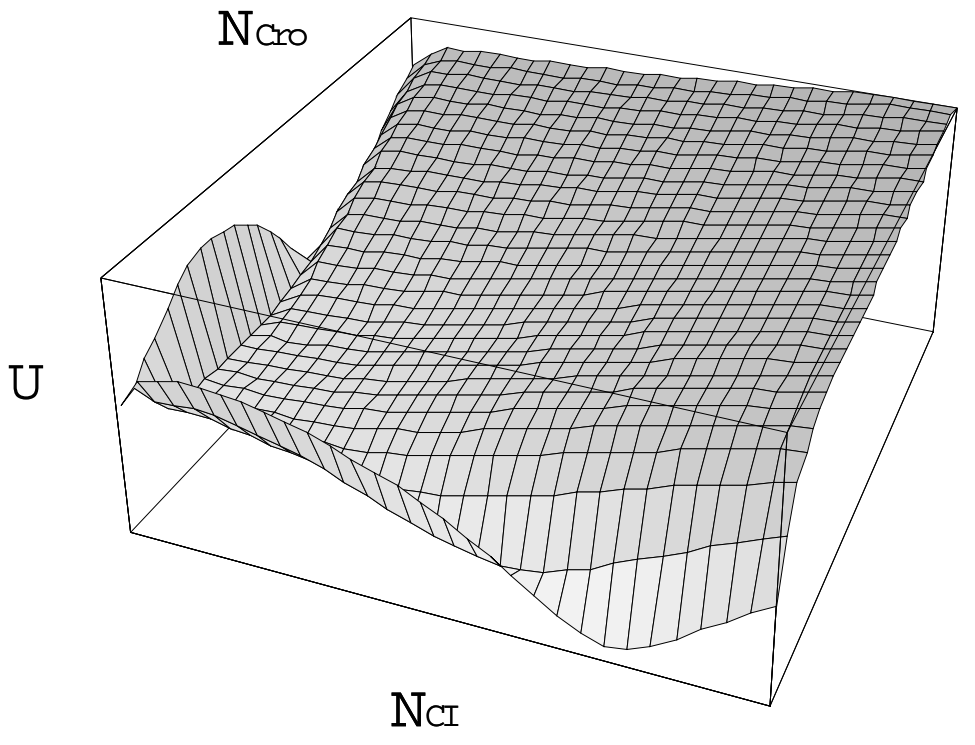

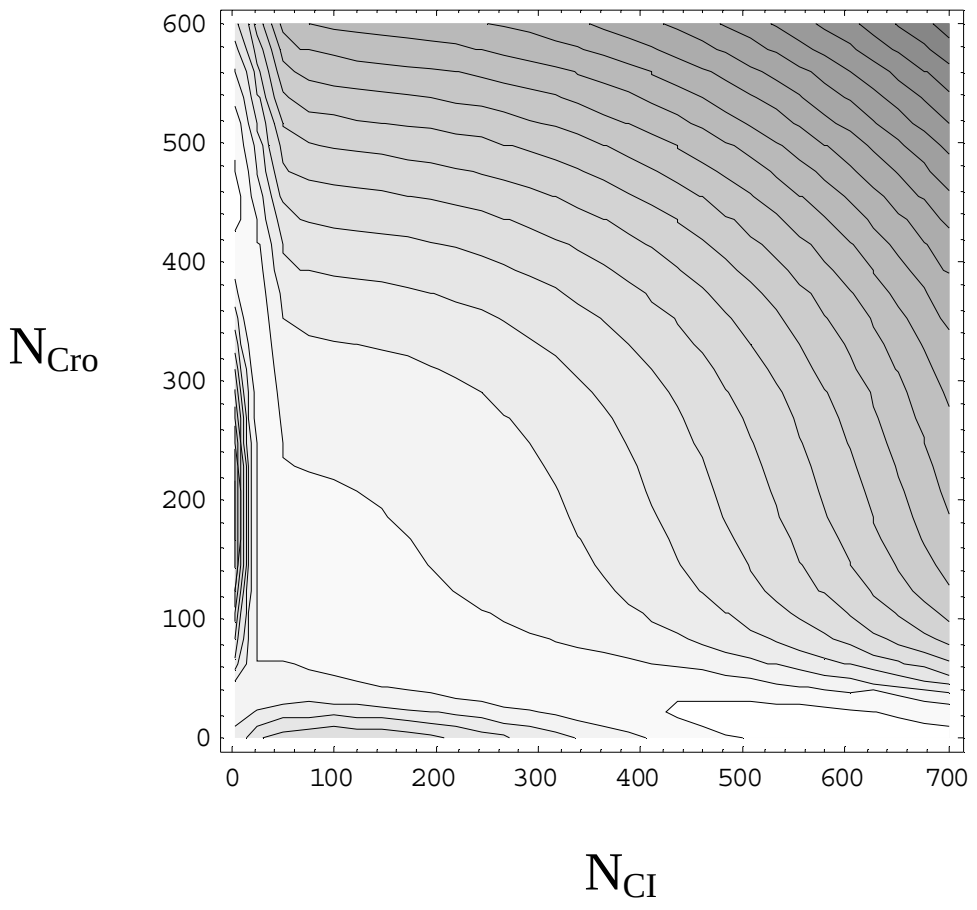

Fig. 4. The potential U of wild type phage plotted on logarithmic scale (a) and as a contour map (b). There are two potential minima corresponding to the lysogenic and lytic states. Connecting these two states is a narrow potential valley. The highest point along this valley is the saddle point. The most probable state of the phage is at either of the potential minima. The fluctuation may bring the phage from the original potential minimum, moving along the valley and across the saddle point to reach another potential minimum. The rate for such a switching event is given by the Kramers rate formulae.

The antisymetric matrix $\Omega$ may be represented by a single scalar B along the $z$ direction $\Omega\mathbf{F}$ = B $z\times$ $\mathbf{F}$, assuming $x$= $N_{CI}$, $y$=$N_{Cro}$. The transverse field B for the wild type is obtained by numerically solving Eq.(13) and the results are plotted in Fig. 5. This field is small except at the region along two axes. Along the two axes, the transverse B field has no effect, since the motion is guided by the steep potential to a valley. Once the phage evolves away from origin when both CI and Cro number are small, in the later development the transverse force may be taken out of Eq.(10) without changing the dynamics of the phage. In both the calculation of the relaxation time and the lifetime of lysogenic state, we may ignore the transverse force for the above reason.

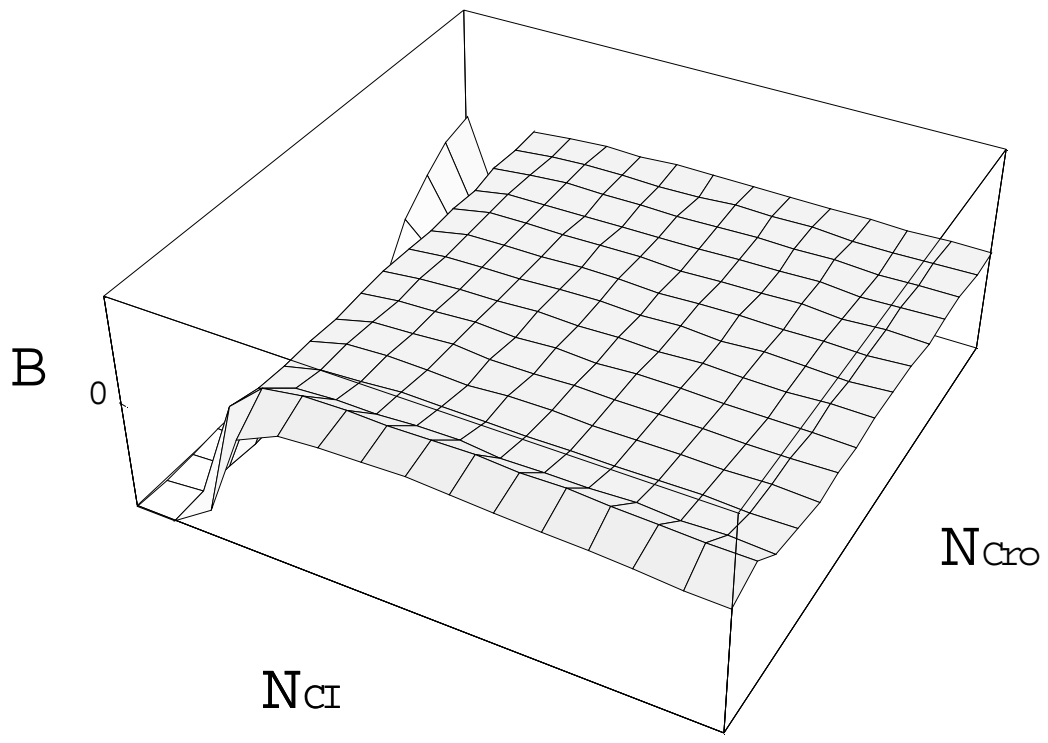

Fig. 5. Plot of the only matrix element B of the antisymmetric matrix $\Omega$. Except for the regions along either of the axes, B is practically zero.

### V. 3 Protein numbers for lysogenic and lytic states

The positions of the potential minima give the average protein number for lytic and lysogenic states. In lysogenic (lytic) state, practically only CI (Cro) proteins exist. By varying the parameters in the model, we may also obtain information on how the protein numbers change when the experimental condition varies (shown in Table II). When the temperature increases, the protein numbers in both lysogenic and lytic state increase. If the protein degradation time of CI decreases, CI protein numbers in lysogenic state decreases. Such a decrease, as we will show later, decreases the stability of the lysogenic state, tipping the phage towards lytic growth.

| | Relative CI level in lysogen | Relative Cro level in lysis |
|---|---|---|
| RT=0.607 | 90% | 80% |
| RT=0.627 | 110% | 130% |
| $\tau_{CI}/2$ | 63% | 100% |
| $\tau_{Cro}/2$ | 100% | 87% |
| $2S_{CI}$ | 160% | 100% |
| $2S_{Cro}$ | 100% | 120% |

Table II. Calculated relative protein numbers in lysogenic and lytic states for variations in temperatures (first two rows), for decreased protein life times (3rd and 4th rows) and for increased transcription rates (5th and 6th rows) respectively.

### V.4 Protein distribution for lysogenic state

The potential near its minimum gives the protein distribution around the average value. Such a distribution may be measured experimentally, although it has not been done for phage λ. The most likely measurement is the CI protein distribution in a lysogenic culture. We give the shape of the distribution in Fig. 6, calculated using Eq.(15). The shallower potential confinement of mutant results in a wider spread of protein number.

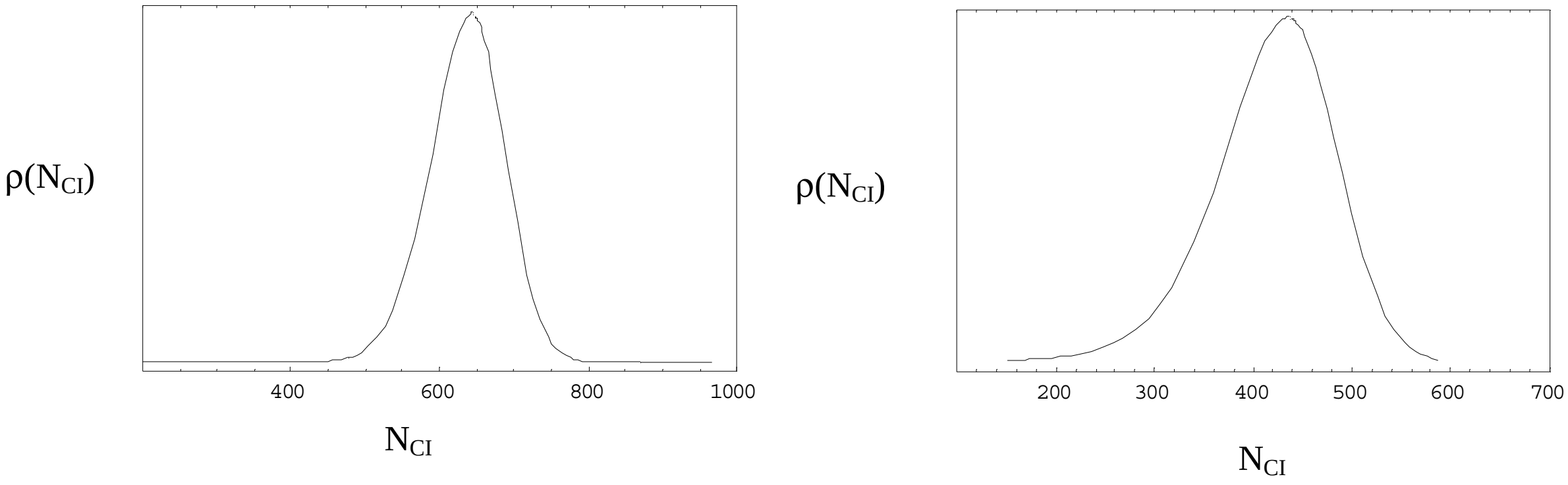

Fig. 6. CI Protein distribution of wild type (left) and mutant $\lambda O_R323$ (right), according to Eq.(15).

### V.5 Relaxation time for lysogenic state

When the protein numbers deviate from their value at the potential minimum after a perturbation, they tend to relax to the value at the potential minimum (shown in Fig. 7). The relaxation time is given by the potential near its minimum and the strength of friction: $\tau = \eta/U''$, here $\eta$ is the strength of friction, $U''$ the second derivative of potential, both $\eta$ and $U''$ along the direction of perturbation. We give the relaxation time for the wild type and mutants measured in Little *et al.*'s data in Table III.

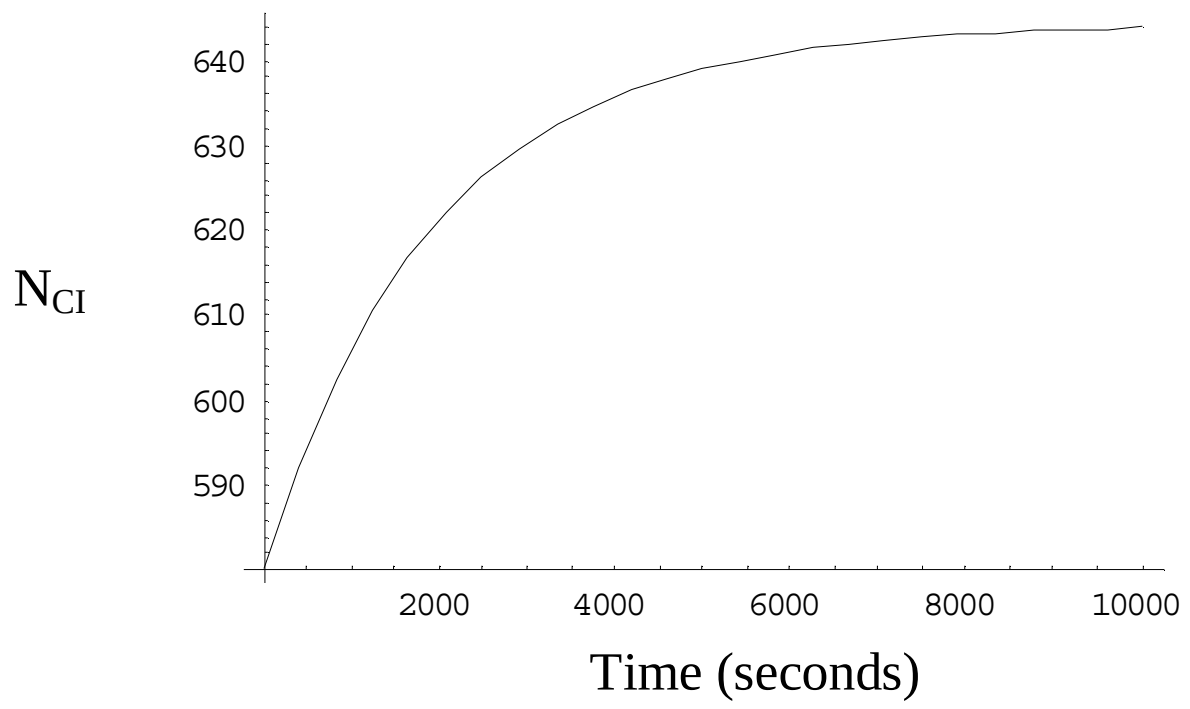

Fig. 7. The relaxation of CI protein numbers after a perturbation.

| Phage | Relaxation time of Lysogenic State (seconds) | Relaxation time of Lytic State (seconds) |
|---|---|---|
| $\lambda^{+}$: | $2\times10^3$ | $1\times10^3$ |
| $\lambda O_R121$ | $4\times10^3$ | $1\times10^3$ |
| $\lambda O_R323$ | $3\times10^3$ | $2\times10^3$ |
| $\lambda O_R3'23'$ | $2\times10^3$ | $1\times10^3$ |

Table III. Relaxation times for lysogenic and lytic states.

## V.6 Stability of lysogenic and lytic states

First we discuss the intrinsic stability of the lysogenic state when *recA* is disabled. When the phage is grown in lysogen culture, occasionally fluctuation may bring the phage λ to lytic growth. The potential difference needed to calculate the Kramers rate of escape is known after we solve for U. We still need the attempt frequency. The calculation of attempt frequency is usually involved and it has a less dramatic influence on the escape rate than the energy difference. Therefore we simply treat it as a fitting parameter, which gives $\omega_0 = 4.0\times10^{-4}$ / minute. The potential energy difference for the mutants in Little *et al.*'s experiments are given below, Table IV. For the sake of completeness, we also include the lifetime of lytic states. This is the time scale that a phage grows in lytic state may change to lysogen state. This time scale is not observable of wild type phage since phage lyze the cell after one hour or so. The intrinsic lifetime of lytic state can only be observed if the lysis is suppressed[43,44].

| Phage | Potential Difference U(Lysogen) –U(Saddle) | Lifetime of Lysogen State (minutes) | Potential Difference U(Lysis)-U(Saddle) | Lifetime of Lysis State (minutes) |
|---|---|---|---|---|
| $\lambda^{+}$: | $\Delta U = -7.49$ | $4.5\times10^6$ | $\Delta U = -8.15$ | $8.7\times10^6$ |
| $\lambda O_R121$ | $\Delta U = -5.14$ | $4.2\times10^5$ | $\Delta U = -1.29$ | $9.1\times10^3$ |
| $\lambda O_R323$ | $\Delta U = -1.67$ | $1.3\times10^4$ | $\Delta U = -11.2$ | $1.8\times10^8$ |
| $\lambda O_R3'23'$ | $\Delta U = -6.63$ | $1.9\times10^6$ | $\Delta U = -2.35$ | $2.6\times10^4$ |

Table IV. Potential barrier heights between potential minima and saddle points for lysogenic and lytic states, and lifetime for these states.

For the r*ecA*$^+$ phage, CI monomers may be cleaved by RecA protein. When the phage is not provoked and the SOS is not activated, it is reasonable to assume that such a cleavage of CI monomer only happens occasionally. The major influence to the phage is to increase its noise level to CI proteins. We find that if the noise level is doubled the potential barrier height is lowered in half. The fraction of lysogens that has switched to lysis for such a noise level agrees with the experimental results well (shown in Table II).

## V.7 Robustness of the genetic switch

The robustness of the genetic switch is understood as its insensitivity to certain parameter variations and its tolerance to certain genetic polymorphism. The stability of the λ switch against mutation is measured by Little *et al.*[15] and summarized in Table II together with our calculation results. Such a robustness was not reproduced by the calculation by Aurell *et al.*[16]. They found that it was impossible to reproduce simultaneously the measured stability of wild type *recA*⁻ lysogens and the existence of stable lysogens in the mutant 323.

We run a straightforward test of the switch based on the new set of parameter values. Since the existence of two potential minima and one saddle point indicates the stability of the switch, we look for the range for configurations. We find that for the average CI monomer number variations from $2\times10^{-2}$ to $1\times10^{2}$ relative to the wild type value, by substituting $N_{CI}$ in Eq.(8) with $\alpha N_{CI}$, and Cro monomer number from $5\times10^{-4}$ to $1\times10^{4}$ relative to the wild type value, the switch is stable. We also let the CI degradation time $\tau_{CI}$ vary from 0.1 to 15 relative to its wild type value and $\tau_{Cro}$ from $1\times10^{-2}$ to $5\times10^{2}$ relative to its wild type value, again the switch is stable. Note that one of parameters has been varied over 9 orders of magnitude. With all above tests on the stability, we conclude that this phage λ switch is remarkably stable, insensitive to small changes in parameters (Fig.8).

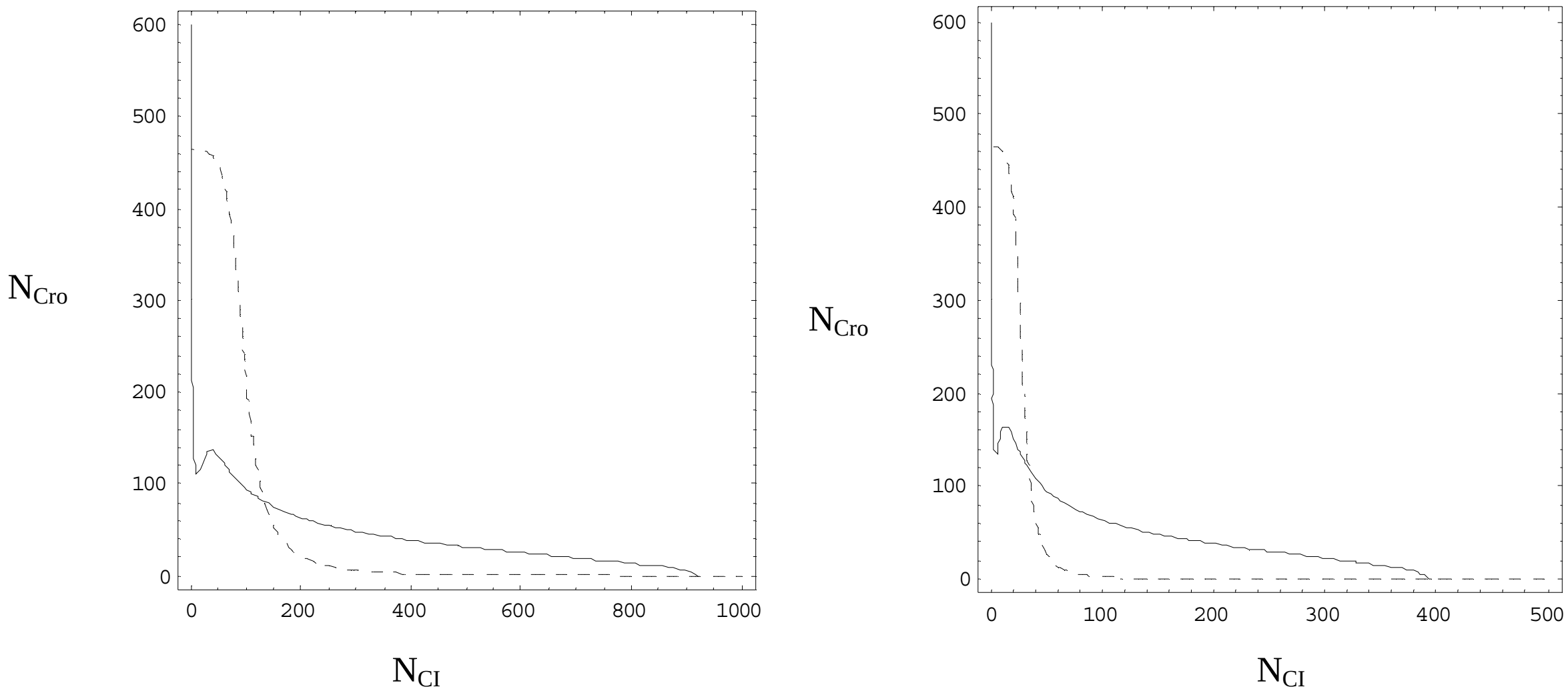

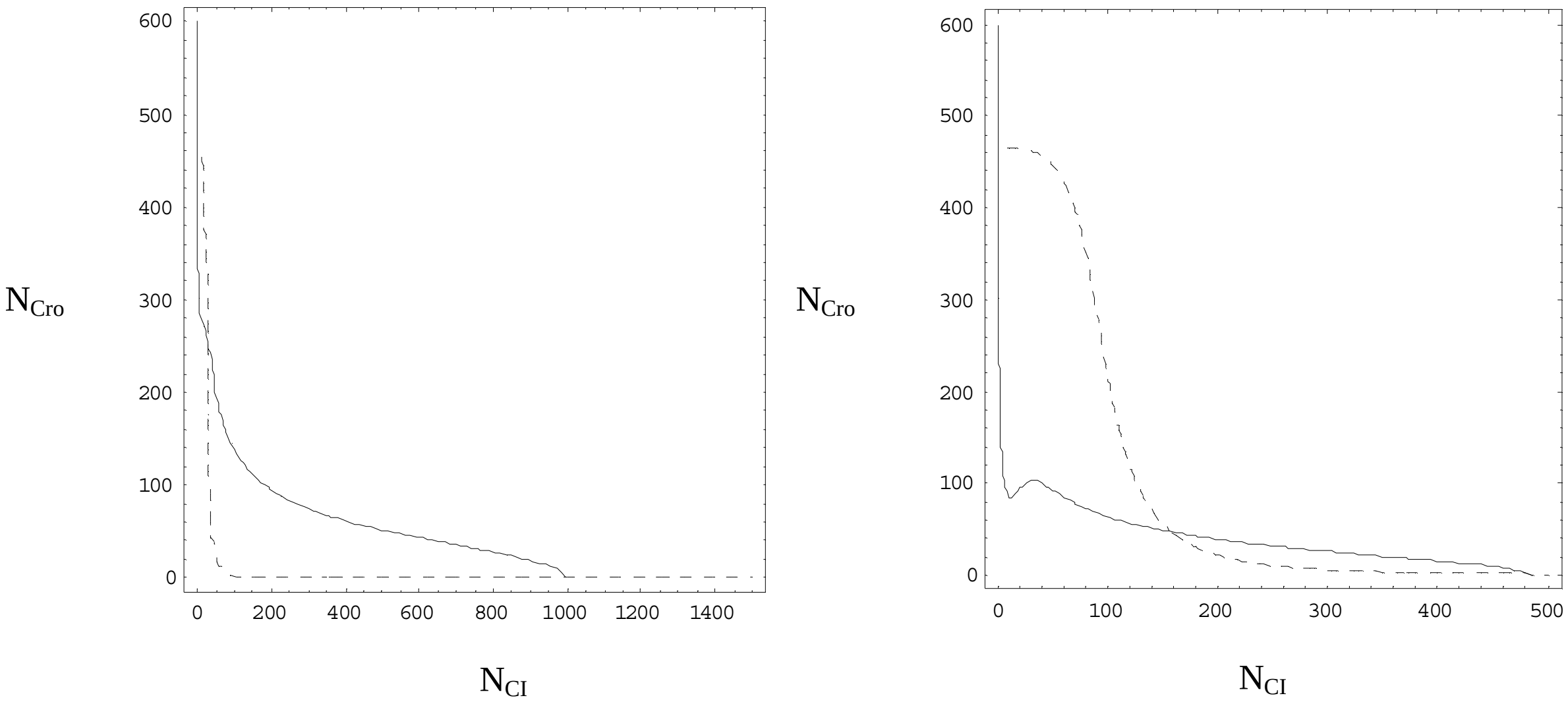

Fig.8. Demonstration of the robustness of the wild type phage genetic switch against parameter variations. Lines of $d\langle N_{CI}\rangle/dt = 0$ (solid) and $d\langle N_{Cro}\rangle/dt = 0$ (dashed) for: (a) reduced CI protein number by replace CI number $N_{CI}$ by $0.1N_{CI}$; (b) $\tau_{CI}$ is reduced to half; (c) CI transcription is doubled; (d) $\tau_{CI}$ is reduced to half and CI protein number is reduced by replacing CI number $N_{CI}$ with $0.1N_{CI}$. For all these cases, there are three intersections giving two potential minima and one saddle point. Therefore they all have switch structure.

By systematically changing the binding energies and other parameter, we try to probe the key element responsible for the robustness of the switch. We find that the cooperative binding of CI dimers is the most important factor. If we reduce the cooperative binding of CI dimers by half, the wild type is not affected significantly. However, the mutant $\lambda O_R323$, which is observed to be stable, becomes unstable in the calculation (shown in Fig .9). We may understand the effect of the cooperative binding as a method to enhance positive feedback. With this cooperativity, a positive feedback is turned on when CI dimers bind to either $O_{R1}$ or $O_{R2}$.

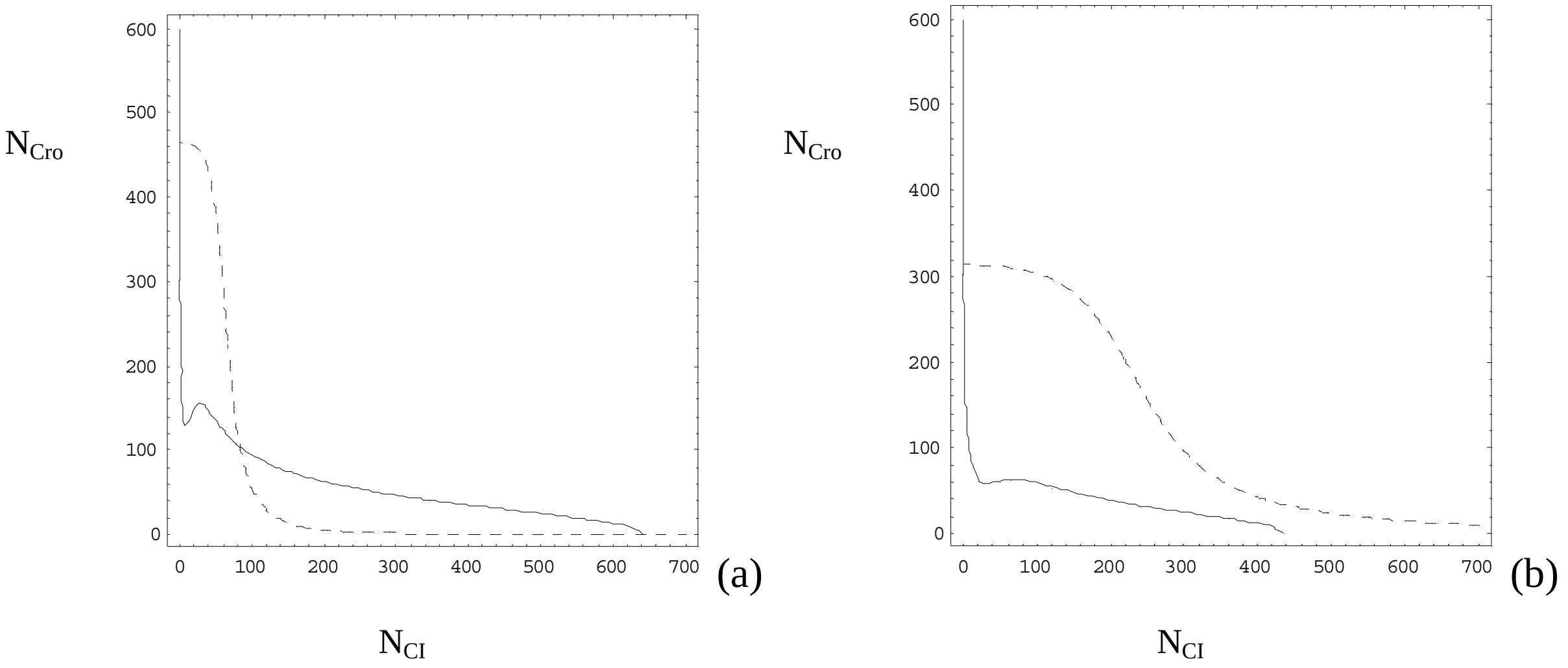

Fig.9. If the cooperative binding energy of CI dimers is reduced by half, the wild type still has a switch structure shown in Fig .9(a), by lines of $d\langle N_{CI}\rangle/dt = 0$ (solid) and $d\langle N_{Cro}\rangle/dt = 0$ (dashed). However, the mutant $\lambda O_R323$, shown in Fig. 9 (b) shows only one fixed point corresponding to lytic growth.

## V.8 Switch efficiency

The analysis of robustness of phage λ genetic switch demonstrates that its epigenetic states are robust against parameter changes, including protein number change. Then how does the switching take place? From theoretical point of view, there are two channels that the phage can be induced from lysogenic growth to lytic growth. In reality phage seems to use both of these strategies. For clarity, we begin by discussing these two channels separately.

The first channel of induction is to increase the noise level of CI protein number while keeping all the other conditions intact. Mathematically it means to increase $\zeta_{CI}$ in Eq.(8) and $D_{CI}$ in Eq.(9), while keeping all the other terms in Eq.(8) and Eq.(9) unchanged. The friction matrix $\Lambda$ is changed through the decomposition procedure. As a result, the potential energy U is also changed. Therefore for different noise level, the phage moves in different potential landscape. Such a change of noise level has a drastic effect. It changes the minima of the potential well of lysogen by making it shallower. As a good approximation, the barrier height of the lysogen potential well scales inversely with the noise strength. Doubling the noise level reduces the potential barrier by half. As a result, the increased noise level drastically decreases the lifetime of lysogenic state, as shown in Fig. 10. The lifetime of the lytic state, on the other hand, remains unchanged. The combination of these two changes in the potential landscape brings the phage to lytic growth efficiently.

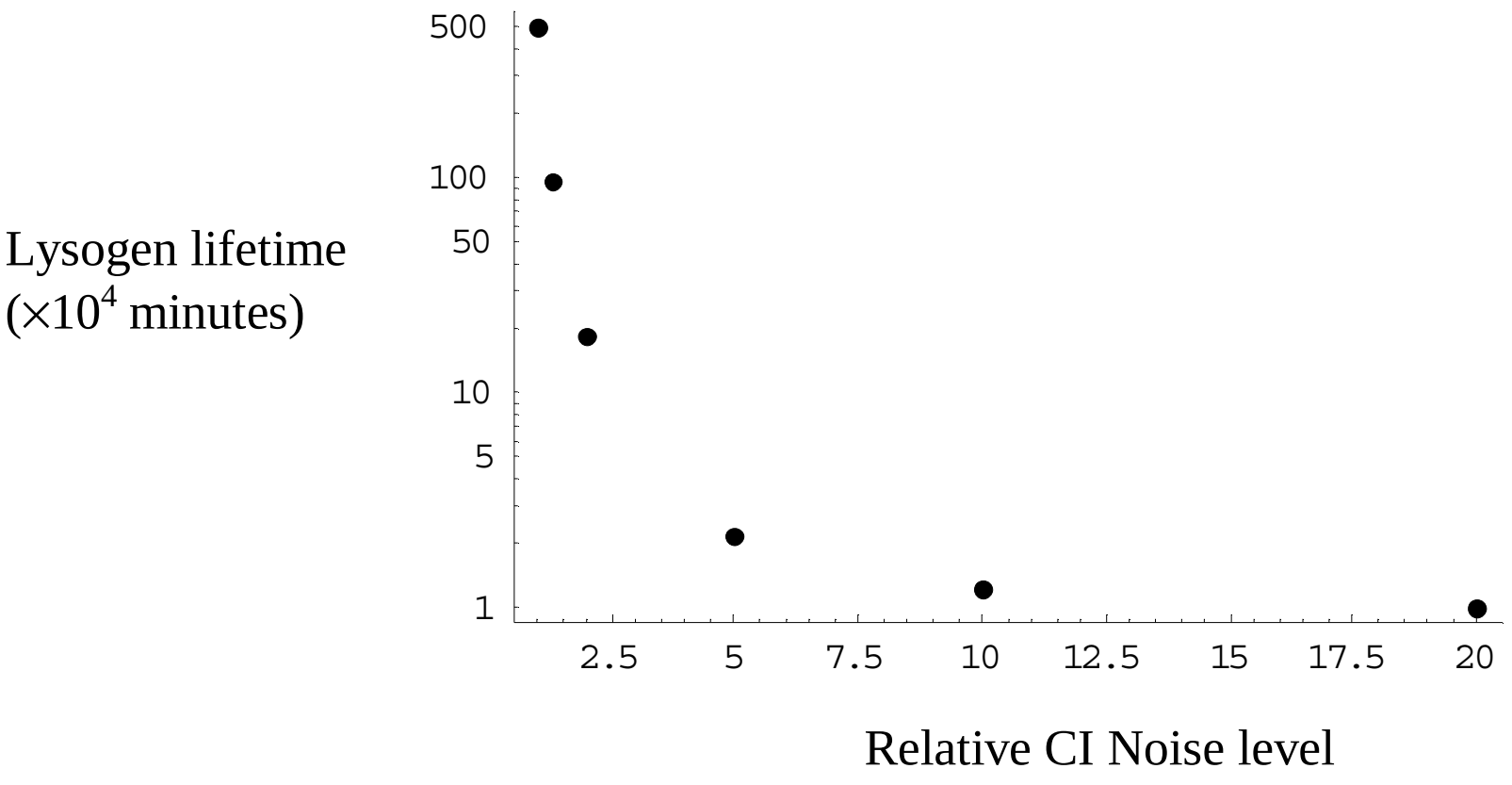

Fig 10. Lysogen lifetime when the noise level for CI proteins increases. If CI noise level doubles, the lysogen lifetime decrease for more than 2 orders of magnitude. It reflects that the noise level controls the lysogen lifetime through an exponential relation.

The second channel is through the deterministic terms in Eq.(8). For the deterministic terms, for example, introducing CI monomer cleavage is equivalent to substitute $N_{CI}$ in Eq.(8) with $\alpha N_{CI}$, here $\alpha$ is a factor represent cleavage strength (Fig. 11). If $\alpha$ is smaller than 0.02, we find that lysogenic state is no longer stable, i.e. no longer a potential minima. The interpretation of such a small $\alpha$ is that almost each of the CI monomer is cleaved. If $\alpha$ is small, say 0.1, meaning 90% CI monomers are cleaved, the lysogenic state is still stable with a lifetime almost unchanged. Apparently a uniform CI cleavage alone without introducing extra noise to CI levels is not an efficient way for induction.

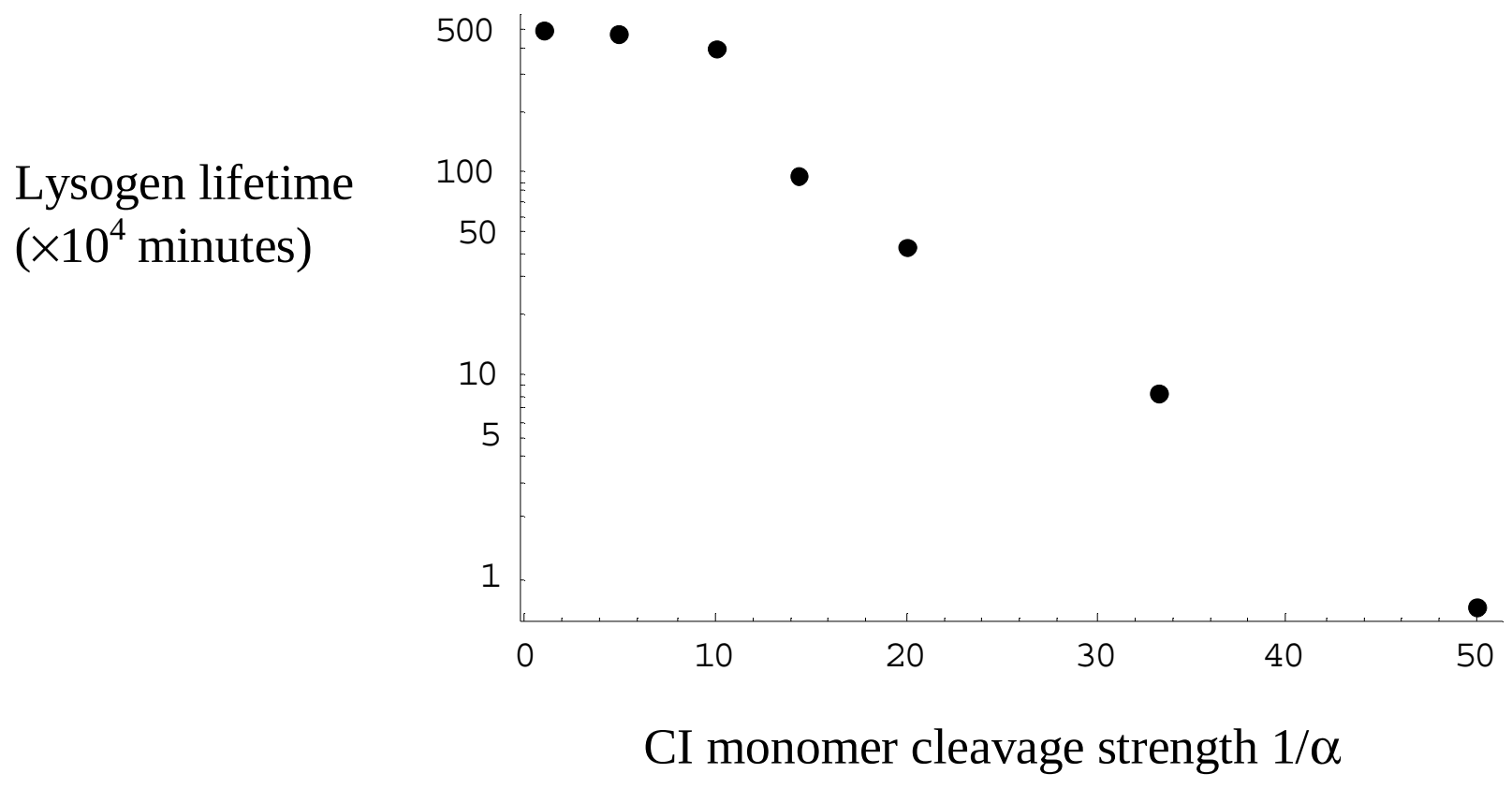

Fig 11. Lysogen lifetime when CI monomer is cleaved by RecA. The cleavage strength is represented by $1/\alpha$. Here $\alpha$ is the relative CI monomers survived the cleavage. If 90% of CI monomers are cleaved, $1/\alpha$ =10, the lysogen lifetime remains almost unchanged. Fig. 10 and Fig. 11 indicate that the noise level increase due to RecA cleavage plays an important role in switching.

Phage may have used both of these two channels. The second channel is obviously used, since RecA cleaves CI monomers. The strong indication that the first channel is also used comes from the observations that without provocation, the *recA*$^+$ phage shows a much shorter lifetime for the lysogenic state compared to *recA*$^-$ phage. Such a significant reduction of lysogen lifetime without activating RecA proteins on an observable scale can be explained by doubling of the CI noise level. Fig.12 gives schematic explanations of the switching process.

In the early work by Shea and Ackers[6], stochastic effect was not included. In their model even a shallow lysogenic potential minimum would confine the phage to continue growing in lysogenic state. Switching happens only when the lysogen potential minimum disappears completely. For the parameters they used, they find that when 20% of CI monomers were cleaved, such a switching would happen. As pointed out by Aurell *et al.*[16], in these early works the genetic switch modeling results do not show the observed robustness. After we require that the genetic switch should demonstrate the observed robustness, the disappearance of the lysogenic potential minimum is pushed down to 2%. However, the actually switching happens before the disappearance of lysogen potential minimum when the lysogen potential minimum is too shallow to confine the fluctuations. If 10% of CI monomer is cleaved, at least we expect a 10-fold increase in $D_{CI}$ due to the reduced CI monomer numbers. The potential barrier for lysogenic state reduces to less than 1, therefore becomes too shallow to allow continual lysogenic growth.

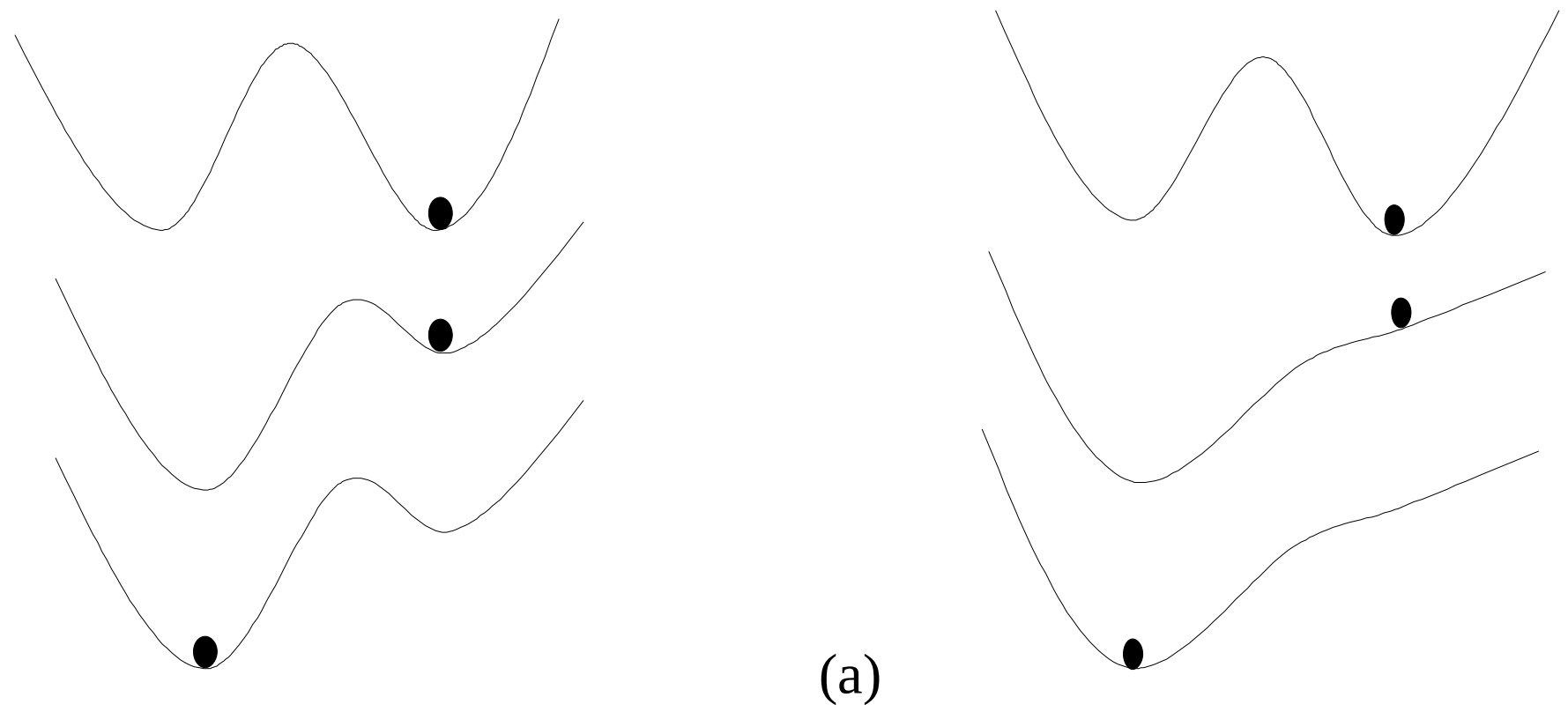

Fig 12. (a) Illustration of the switching mechanism from the current work. Before switching the phage grows in lysgoenic state. The potential barrier separating the lysogenic state and the lytic state is high. When *recA* is activated, this barrier is lowered. The lifetime of lysogenic state reduces drastically and the phage switches to lytic state. (b) Switching mechanism of Shea and Ackers[6]. In their work fluctuation was not included. Switching was only possible when the lysogenic state is no longer a potential minimum. When stochastic effect is included, the switching happens when the lysogenic potential minimum become too shallow to confine fluctuation.

## V.9 Quantitative comparison with experiment of Little *et al.*

We summarize the calculation results related to measurement of Little *et al.*[15] in Table V since their data are most update and systematic. In their experiment, they measured the free phage per lysogenic cell for both $recA^+$ and $recA^-$ phage, but did not convert the $recA^+$ into fraction of lysogens that switched to lytic state. If we assume the burst size for both the $recA^+$ and the $recA^-$ phage are similar, our calculation for the $RecA^+$ protein agrees with their measurements quantitatively.

| Phage | Relative CI level in lysogen | Relative Cro level in lysis | Switching frequency to lytic state ($recA^-$) per minute | Switching frequency to lytic state ($recA^+$) per minute |
|---|---|---|---|---|
| | Theoretical (experimental) | Theoretical | Theoretical (experimental*)) | Theoretical |
| $\lambda^+$ | 100% (100%) | 100% | $1\times10^{-9}$ ($2\times10^{-9}$) | $1\times10^{-5}$ |
| $\lambda O_R121$ | 20% (25-30%) | 100% | $3\times10^{-6}$ ($3\times10^{-6}$) | $3\times10^{-5}$ |
| $\lambda O_R323$ | 70% (60-75%) | 70% | $7\times10^{-5}$ ($2\times10^{-5}$) | $1\times10^{-4}$ |
| $\lambda O_R3'23'$ | 50% (50-60%) | 130% | $1\times10^{-7}$ ($5\times10^{-7}$) | $2\times10^{-5}$ |

Table V. Comparison between the calculation and the experiment data (in parentheses) by Little *et al*.[15]. Here *) indicates that the estimated wild type data from Little[17] is used. The wild type biological data were used to find out the difference between *in vivo* and *in vitro* molecular parameters, as listed in Table I. The relative CI level and switch rate of $\lambda O_R121$ were used to fine tune parameters. Rest theoretical entries are then calculated directly from our model.

In Table V the bi-stability of the gene switch in phage $\lambda$ is assumed, and the protein levels in the lytic state are calculated. This is of course not the case for the wild type, hence posts a question to test the calculated Cro level experimentally. One way to realize the bi-stability may be by suppressing the lyses, achieving the so-called anti-immune phenotype[43,44].

As discussed in the formulation of the present model, we have made the simplified assumption of treating all chance or probability events as Gaussian white noise. This assumption affects two testable biological quantities: the lifetime of lysogenic state, Eq.(16), and the shape of CI number distribution in lysogenic state, Eq.(15) and Fig.6. Simultaneous measuring both of them can be used as a consistent check to the Gaussian white noise assumption. We have treated the effect of $recA^+$ to switching dynamics as that of a Gaussian white noise to simplify our calculation, in the same reasoning of minimal modeling type approach in the present paper. In fact, we have assumed that with $recA^+$ the total noise strength doubles, compatible with discussions presented by Fig. 10 and 11. Using this assumption we calculated the lytic switching rates, represented by the last column of Table V. The CI distribution with $recA^+$ should be twice as broad as in the case with $recA^-$. Both results are subjected to the experimental test.

There may be some chance events that cannot be treated as Gaussian white noise in the present formulation. One example has been already suggested in biological experiments[17], the $p_{RM}$240 mutation which greatly weakens the promoter therefore the ability to produce CI. This mutation makes the lysogens barely stable, and is estimated to responsible for at least 99% of observed lytic switching in the wild type. We have used this input for both Table I and V. We could, however, re-calculate the switching rates to lytic state of all strands, assuming the same minimal model, with the same forms of functions for the switching rate, but with the previous experimental data[15]. The switching rates obtained in this way are: wild type($\lambda^+$), $2\times10^{-7}$; $\lambda O_R$121, $\lambda O_R 2\times10^{-6}$; $\lambda O_R$323, $7\times10^{-5}$; $\lambda O_R$3'23', $5\times10^{-7}$. Indeed, the stability of the wild type decreases by more than 2 orders of magnitude. The overall noise strength is increased by 60% for the wild type, resulting a broader CI distribution in lysogenic state. There is no appreciable change in other quantities, such as the protein level. The only noticeable overall change in molecular parameters is the *in vivo* cooperative energy, from –6.4 kcal/mol to –6.7 kcal/mol. A good overall quantitative agreement exists between modeling and experiment.

It is a fact that any mathematical modeling in natural science should have empirical input to completely fix its mathematical structure. For the modeling of phage λ there is an already large body of molecular data which enables us to nearly pin down our model. The additional freedom in our parameters is fixed by data from wild type, such as the switching frequency. Above less-than-expected sensitivity of our mathematical structure to this frequency that a few percentage of change in molecular parameters can result in two orders of magnitudes change in frequency is a remarkable demonstration of the internal consistence of our modeling. It demonstrates that the switching is exponentially sensitive to some molecular parameters. In addition to more theoretical effort to go beyond our present minimal modeling, it is clear that more experiments are needed in this direction to test the present model: The precise *in vivo* molecular parameters and the distributions and time-correlation of protein numbers in our model should be viewed as predictions.

## V.11 Determining dynamical structure from direct experiments

We have introduced four dynamical quantities for a gene regulatory network: friction, potential, the transverse force and the stochastic force. The friction and the strength of the stochastic force are related. For the genetic switch, the transverse force is irrelevant to the dynamic properties. Therefore the two crucial quantities for a genetic switch are the friction and the potential. Those quantities can be calculated from the more microscopic modeling with molecular parameters. The present quantitative success lies in the allowance of the *in vivo* and *in vitro* differences, and of various noise contributions. However, those four quantities may be directly determined biologically.

There are three experiments which allow to determine the friction and the potential for a genetic switch (Fig.13). The first one is the protein distribution around each of the epigenetic states given by

$$\rho(\mathbf{N}) = \rho_0 \exp(-U(\mathbf{N})), \qquad \text{Eq.(18)}$$

here $\rho_0$ is a normalization constant. The protein distribution is explicitly measurable. Once $\rho(\mathbf{N})$ is measured, the potential $U(\mathbf{N})$ near the potential minima can be determined from it: $U(\mathbf{N}) = -\ln(\rho(\mathbf{N})) + \ln(\rho_0)$.

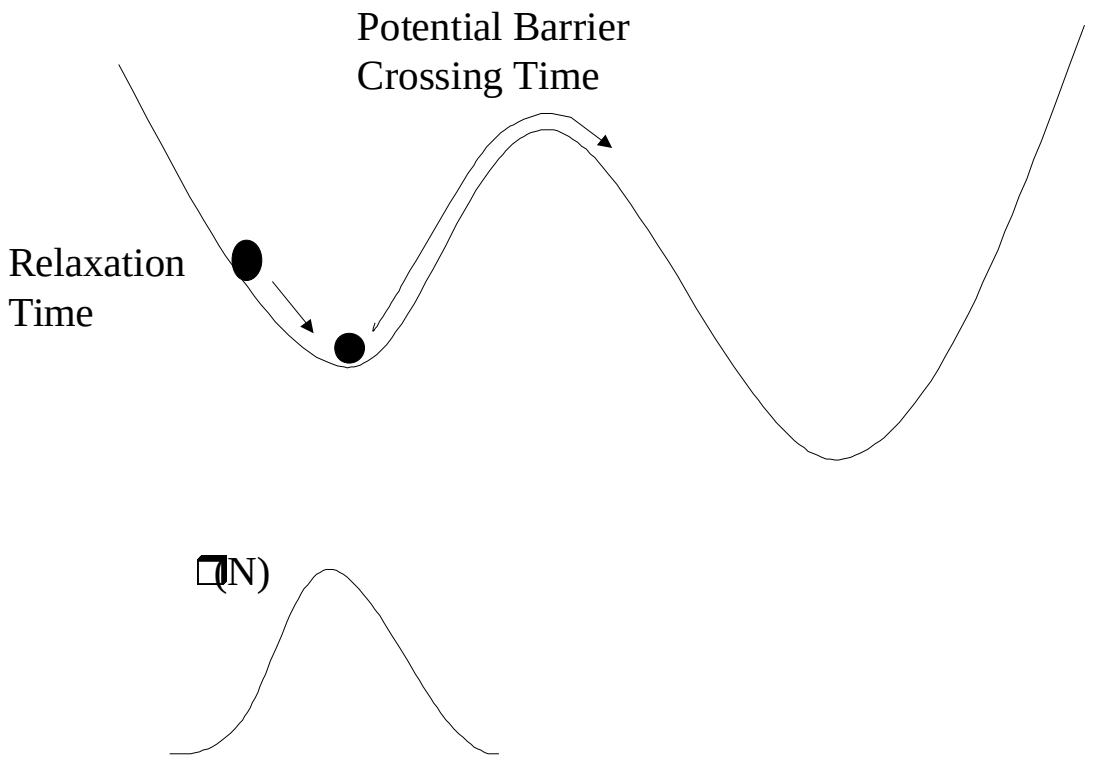

Fig 13. Three types of experiments directly probe the dynamical structure of a genetic switch and determine the dynamic components. The measurement of protein number distribution determines the potential function at each of the epigenetic state. The additional information on the relaxation time determines the strength of friction. The lifetime of each epigenetic state determines the height and shape of potential barrier.

The second experiment is the relaxation time. It is determined by both the potential near the minimum and the friction,

$$\tau = \eta / U'', \qquad \text{Eq.(19)}$$

here $\eta$ is the strength of friction, which gives the friction matrix $\Lambda$ along the path of relaxation. $U''$ the second derivative of potential. Since potential can be obtained from $\rho(\mathbf{N})$, relaxation time can be used to obtain friction: $\eta = \tau U''$.

The third experiment is the lifetime of epigenetic states. The probability of phage evolving from one epigenetic state of growth to another is given by the Kramers rate formulae, our Eq.(16),

$$P = \omega_0 \exp(- \Delta U_b) ,$$

where $\Delta U_b$ is barrier height and $\omega_0$ is the attempt frequency. $\omega_0$ is given by the friction and the curvature of the potential barrier. The curvature of the potential is related to the height of the potential barrier and the shape of potential near its minimum. Therefore $\Delta U_b$ can be determined from the lifetime of its epigenetic state: $\Delta U_b = \ln(\omega_0) - \ln(P)$.

### V.11 Dynamical structure analysis as a phenomenological description

The gene regulatory network for phage λ is a complex dynamical system. It took decades of ingenious experimental research and laborious work to collect parameters needed for this mathematical modeling. For a more complicated system, resource and time may limit the ability to study each molecular element in such a great detail. A method that is less demanding on the details yet still can capture the main features is of great interest. The dynamical structure theory analysis provides a guidance to build such a phenomenological model.

The quantities introduced in dynamical structure theory, the friction, the potential gradient, the transverse force and the stochastic force associating with the friction are all measurable quantities at the given description level. These are quantities similar to temperature, pressure, free energy in thermodynamics, which are determined by the microscopic details but can be measured independent of the details. Once the relationship between these quantities are established, as shown in Eq.(10), we are ready to write down the effective equation of motion for the network without resorting to details.

## VI. Perspective

Ten years ago, forty year after the discovery of dramatic switching behavior of bacteriophage λ, it was acknowledged[2] that we should now "appreciate the insight of those earlier workers who recognized in the growth of this virus — in particular its ability to grow in two different modes — a revealing example of gene expression". The importance of these studies lies in its context, that the molecular mechanism underlying gene expression apply to many other biological regulatory processes as well.

A comparison may be drawn between the study of phage λ in molecular biology and the study of hydrogen atom in modern physics. In both cases, the objects under study are the simplest of their kinds: The phage λ one of the simplest living organism and hydrogen atom the simplest atom. In the process of understanding each of them, new theoretical frameworks and experimental methods emerge. Quantum mechanics, which developed associating with the study of hydrogen atom, has profoundly changed the world of science and technology. Although we cannot foresee what eventually the study of phage λ will bring us to, there is no doubt of its importance.

We believe that present study is an example demonstrating the power of quantitative approach. It shows that mathematical framework can serve as both quantitative description and biological discovery tools. The biological properties under study, the robustness of the phage genetic switch, the stability and efficiency of the switch and the mechanism that allows for the coexistence of them, are discussed through mathematical modeling. By changing the parameters in the model and study the outcome, we find the importance of cooperativity in the robustness of genetic switch and the roles of fluctuation in the efficiency of switching. The gene regulatory system in a living organism is a complex network. Its dynamical equation involves at least tens of molecular parameters. These parameters may be determined experimentally to a reasonable degree of accuracy. As we show in this work, even for organisms as simple as phage λ for which the tens of molecular parameters have been indeed measured, we may still allow differences between *in vitro* measured parameters and *in vivo* ones. Therefore for an organism more complicated than phage λ, carrying a full scale modeling with all measured molecular parameters appears formidable. For these complicated systems, the last part of our discussion in previous section, to build a phenomenological model from direct experiments, may offer a practical help. It is then obviously important to carry out a full-scaled modeling as much as we can for simple systems like phage λ to gain insight in order to build phenomenological models for more complicated systems.

In spite of the present quantitative success, the study of gene regulatory network is only at its beginning. We have not yet touched deep biological questions, such as why the phage chooses such a structure or what are the evolution principles guided this choice[45]. Most importantly, even after we acquire a thorough quantitative description of the phage λ gene regulatory network, we still need to find out how to improve the knowledge and method we would have developed in order to apply to more intricate higher organisms. Once those questions are answered, it is not inconceivable that the phage λ will lead us to an understanding much deeper and grander than what we can imagine right now.

## Acknowledgement

We thank L. Hartwell for emphasizing the importance of biological modeling with incomplete molecular parameters and J.W. Little for critical comments and for sharing his unpublished data. We also thank G.K. Ackers, H. Eisen, D. Galas, M.C. Mossing, V. Ng, and M. Olson for numerous valuable comments. This work was supported in part by the Institute for Systems Biology (P.A. and L.H.) and by a USA NIH grant under HG002894-01 (P.A.), and by a USA NSF grant under DMR 0201948 (L.Y.).